\documentclass[12pt]{article}

\usepackage{rotating}

\usepackage{multirow}
\usepackage{amsmath,amssymb,graphicx, graphics, color,verbatim} %use drftcite in draftmode
\usepackage{cite}
\usepackage{latexsym} 
\usepackage{cancel}
\usepackage{hyperref}

\usepackage{enumerate} 

\usepackage[normalem]{ulem} %for strikethrough

\pdfoutput=1
\usepackage{epsf}
\usepackage{pstricks} 
\usepackage{comment}

\usepackage{array}

\newcommand{\centeron}[2]{{\setbox0=\hbox{#1}\setbox1=\hbox{#2}\ifdim
\wd1>\wd0\kern.5\wd1\kern-.5\wd0\fi \copy0
\kern-.5\wd0\kern-.5\wd1\copy1\ifdim\wd0>\wd1
                                   \kern.5\wd0\kern-.5\wd1\fi}}
\newcommand{\ltap}{\>\centeron{\raise.35ex\hbox{$<$}}
                           {\lower.65ex\hbox{$\sim$}}\>}
\newcommand{\gtap}{\>\centeron{\raise.35ex\hbox{$>$}}
                           {\lower.65ex\hbox{$\sim$}}\>}

\newcommand\ZZ{\hbox{\zfont Z\kern-.4emZ}}
\font\zfont = cmss10 %scaled \magstep1

\newcommand{\cref}[1]{Chapter \ref{c.#1}}

\newcommand{\ba}{\begin{array}}
\newcommand{\ea}{\end{array}}

\newcommand{\beq}{\begin{eqnarray}}% can be used as {equation} or  {eqnarray}
\newcommand{\eeq}{\end{eqnarray}}
\newcommand{\beqs}{\begin{eqnarray*}}
\newcommand{\eeqs}{\end{eqnarray*}}

\newcommand{\bal}{\begin{align}} %this is glorious: arbitrary number of columns, alignment
\newcommand{\eal}{\end{align}}

\def\bi{\begin{itemize}}
\def\ei{\end{itemize}}
\def\ben{\begin{enumerate}}
\def\een{\end{enumerate}}
\def\bc{\begin{center}}
\def\ec{\end{center}}
\def\bt{\begin{table}}
\def\et{\end{table}}
\def\btb{\begin{tabular}}
\def\etb{\end{tabular}}

\def\co{{\mathcal O}}

\def\gev{\, {\rm GeV}}

\def\mass2{mass${}^2$}

  \newcommand{\ww}{
$W^+W^-$
}

\newcommand{\wz}{$W^\pm Z$~}

\newcommand{\ttb}{$t\bar{t}$~}

\newcommand{\zz}{$Z Z$~}

\newcommand{\pt}{$p_T$~}

\begin{document}
\bibliographystyle{unsrt}
\begin{titlepage}
\begin{flushright}
\small{YITP-SB-14-22}
\end{flushright}

\vskip2.5cm
\begin{center}
\vspace*{5mm}
{\huge \bf Transverse momentum resummation  effects in \ww measurements}
\end{center}
\vskip0.2cm

\begin{center}
{Patrick Meade, Harikrishnan Ramani, Mao Zeng}

\end{center}
\vskip 8pt

\begin{center}
{\it C. N. Yang Institute for Theoretical Physics\\ Stony Brook University, Stony Brook, NY 11794.}\\
\vspace*{0.3cm}

\vspace*{0.1cm}

{\tt meade@insti.physics.sunysb.edu, hramani@insti.physics.sunysb.edu,mao.zeng@stonybrook.edu}
\end{center}

\vglue 0.3truecm

\begin{abstract}

The $W^+W^-$ cross section has remained one of the most consistently discrepant channels compared to SM predictions at the LHC, measured by both ATLAS and CMS at 7 and 8 TeV.  Developing a better modeling of this channel is crucial to understanding properties of the Higgs and potential new physics.  In this paper we investigate the effects of NNLL transverse momentum resummation in measuring the $W^+W^-$ cross section.  In the formalism we employ, transverse momentum resummation does not change the total inclusive cross section, but gives a more accurate prediction for the $p_T$ distribution of the diboson system. By re-weighting the $p_T$ distribution of events produced by Monte Carlo generators, we find a systematic shift that decreases the experimental discrepancy with the SM prediction by approximately $3-7\%$ depending on the MC generator and parton shower used.  The primary effect comes from the jet veto cut used by both experiments.  We comment on the connections to jet veto resummation, and other methods the experiments can use to test this effect.  We also discuss the correlation of resummation effects in this channel with other diboson channels.  Ultimately $p_T$ resummation improves the agreement between the SM and experimental measurements for most generators, but does not account for the measured $\sim 20\%$ difference with the SM and further investigations into this channel are needed.

\end{abstract}

\end{titlepage}

%%%%%%%%%%%%%%%%%%%%%%%%%%%%%%%%%%%%%%%%%%%%%%%%%%%%%%%%%%%%%%%%%%%%%%
%%%%%%%%%%%%%%%%%%%%%%%%%%%%%%%%%%%%%%%%%%%%%%%%%%%%%%%%%%%%%%%%%%%%%%% 
\section{Introduction}
\label{s.intro} \setcounter{equation}{0} \setcounter{footnote}{0}
%%%%%%%%%%%%%%%%%%%%%%%%%%%%%%%%%%%%%%%%%%%%%%%%%%%%%%%%%%%%%%%%%%%%%%
%%%%%%%%%%%%%%%%%%%%%%%%%%%%%%%%%%%%%%%%%%%%%%%%%%%%%%%%%%%%%%%%%%%%%
The Standard Model (SM) of particle physics has been tested at a new energy frontier by the Large Hadron Collider (LHC).  SM cross sections were measured at both 7 and 8 TeV, and the SM has passed with flying colors in almost every channel.  Nevertheless there has been one channel that is consistently off at the LHC for both the ATLAS and CMS experiments, the \ww cross section measured in the fully leptonic final state.  This state is naively one of the most straightforward channels to measure both theoretically and experimentally as it is an electroweak final state with two hard leptons.  However, at 7 and 8 TeV ATLAS \cite{atlas7ww,atlas8ww} and CMS\cite{cms7ww,cms8ww} have measured a discrepancy with the SM NLO calculation \cite{Frixione:1993yp,Ohnemus:1991kk} of $\mathcal{O}(20\%)$ and this extends to differential measurements not just simply an overall rescaling.  

This discrepancy is particularly compelling for a number of reasons.  First and foremost, one of the most important channels for the Higgs is the \ww decay channel of which SM \ww is the largest background.  Since this channel doesn't have a particular kinematic feature akin to bumps in the $\gamma\gamma$ or $ZZ$ channels, it is important to understand the shape of the SM background quite well.  CMS \cite{Chatrchyan:2013iaa} and ATLAS\cite{Aad:2013wqa} use data driven techniques to extrapolate and find the signal strength of the Higgs.  While these data driven techniques are validated in many ways, it's often times difficult to find perfectly orthogonal control regions and correlations may arise at higher order in theoretical calculations or because of new physics contributions. Given the shape differences observed, whether or not this is due to an insufficient SM calculation or new physics, It is important to understand that there could possibly be effects which alter the signal strength of the Higgs when the SM \ww channel is understood better \cite{Davatz:2004zg}.  

Another compelling reason for understanding the discrepancy is the possibility of new sub TeV scale physics.  The dilepton + MET final state is an important background to many searches, but even more so, the large $\mathcal{O}$(pb) discrepancy currently observed still allows for the possibility of new $\mathcal{O}(100)$ GeV particles.  While models of this naively would have been ruled out by previous colliders, or other searches at the LHC, in fact it turns out that there could be numerous possibilities for physics at the EW scale.  These include Charginos\cite{Curtin:2012nn}, Sleptons\cite{Curtin:2013gta}, Stops\cite{Rolbiecki:2013fia,Curtin:2014zua,Kim:2014eva} or even more exotic possibilities\cite{Jaiswal:2013xra}. Remarkably, as first shown in\cite{Curtin:2012nn}, not only could new physics be present at the EW scale, it in fact can {\em improve} the fit to data compared to the SM significantly, because it preferentially fills in gaps in the differential distributions when new physics is at the EW scale.  In particular the possibility of particles responsible for naturalness in SUSY being at the weak scale and realizing a solution of the hierarchy problem makes this particularly compelling given all the negative results in other channels.

Finally, It is particularly interesting simply from the point of QCD and the SM to understand why the \ww channel has a persistently discrepant experimental result compared to SM predictions when other similar uncolored final states e.g. $ZZ$ and $WZ$ seem to agree quite well with experiment.   There are potential theoretical reasons within the SM that could explain the difference compared to experiment and to other EW channels.   One of the first points that could be addressed in the context of the \ww measurement is whether or not the fixed order calculation was sufficient to describe the data.  Currently the \ww channel is formally known at NLO, and this is implemented in various NLO MC generators employed by ATLAS and CMS in their analyses.  However, partial NNLO results are also incorporated,  since $gg\rightarrow W^+W^-$ via a quark loop is included through the generator gg2VV\cite{Kauer:2012hd, Kauer:2013qba}. The merging of NLO WW and WWj predictions have been investigated in \cite{Campanario:2013wta, Campanario:2012fk, Cascioli:2013gfa}, while approximate calculations for higher order corrections to $gg\rightarrow W^+ W^-$ are performed in \cite{Bonvini:2013jha}. Theoretically the full NNLO calculation of \ww production turns out to be quite difficult, but within the past year there has been a great deal of progress; the complete NNLO calculation for $ZZ$ total cross section has recently been completed\cite{Cascioli:2014yka}.  The results of\cite{Cascioli:2014yka} are interesting, given that compared to NLO, the NNLO effect can be sizable $\co (10\%)$.  However, when examined closely, if the full NNLO results are compared to the NLO + $gg\rightarrow ZZ$ the difference is less than $\co (5\%)$.  Given this result for $ZZ$, unless there were large differences from a channel with very similar contributions, it would be highly unlikely that the full NNLO result could explain the discrepancies in the \ww result.   

There can be effects beyond the fixed order calculation that matter as well.  As with any calculation there are additional logarithms that arise whenever there is an extra scale in the problem.  For instance threshold resummation logs, or logs of the transverse momentum of the system compared to the hard scale of the system.   These logarithms can either change the overall cross section as in the case of threshold resummation, or the shape of the \pt distribution in \pt resummation.  In~\cite{Dawson:2013lya} the threshold resummation effects were calculated to approximate NNLL accuracy for \ww production, and the effects were found to be small for the overall cross section of $\co (.5-3\%)$ compared to NLO (the NNLO calculation would largely include these logs and thus these effects shouldn't be taken independently in magnitude).  Another contribution which primarily effects the overall cross section, comes from $\pi^2$ resummation\cite{Magnea:1990zb, Ahrens:2008qu, Ahrens:2008nc}.  This has yet to be computed for \ww, however it would affect other EW channels similarly, so the \ww channel shouldn't be singled out and it clearly doesn't explain a discrepancy of  $\mathcal{O}(20\%)$ as measured in that channel.  

While the aforementioned effects primarily affect total cross-section, there are avenues that change the  shape in a differential direction while keeping the total cross section constant. One such effect is $p_T$ resummation first calculated for \ww in\cite{Grazzini:2005vw,Wang:2013qua}.  An interesting difference that arises with \pt resummation, compared to threshold resummation, is the interplay between the effects of resummation and the way the cross section is measured for \ww.  Given that \pt resummation changes the shape of the \pt distribution, and the \pt distribution would be a delta function at 0 at the Born level, QCD effects are crucial for getting this distribution correct.  These effects are normally sufficiently accounted for by using a Parton Shower (matched to LO or NLO fixed order) which only approaches NLL accuracy.  However, in the \ww channel compared to the \wz and \zz channels there is an additional jet veto requirement for the measurement.  This requirement arises because there is an overwhelming background to \ww coming from \ttb production and decay.  The most straightforward way to reduce the \ttb background is to veto on extra jets to isolate the \ww contribution.  Given this jet veto, and the correlation between jet veto efficiency and the \pt shape of the \ww system, there is an added sensitivity to the jet veto and the shape of the \pt distribution that other channels typically don't have. There is precedent for turning to $p_T$ resummation rather than using a parton shower alone when shape differences are important, e.g. the W mass measurement at the D0\cite{D0:2013jba}. 

In this paper we will examine the detailed effects of \pt resummation at approximate NNLL accuracy in combination with how the experimental measurements are performed.  Typically the comparison between \pt resummed processes e.g. Drell-Yan or \zz is done at the unfolded level experimentally.  However, it is the extrapolation from the fiducial cross section to the inclusive cross section that can exactly be the source of a the discrepancy and a new analysis has to be carefully performed to understand the \ww channel.  The difficulty in doing this of course is that in the context of \pt resummation, all radiation is inclusively summed without reference to a jet algorithm, and there is no jet-veto that can be explicitly performed.  In light of this, we undertake a procedure similar to what is done for Higgs production predictions at the LHC using HqT\cite{deFlorian:2011xf} to predict the transverse momentum distribution of the Higgs.  We investigate the effects of taking NLO + parton shower generated events for \ww, reweighting them with the NNLL resummed \pt distribution before cuts, and then applying the cuts to find the fiducial cross section, and how the total cross section should be interpreted.  We find that this leads to $\co (3-7\%)$  changes in the total cross section, for central choices of scales, which reduces the discrepancy.  

A jet veto introduces an additional scale and thus logs related to this scale. Such logs are not identical to the logs accounted for by \pt resummation.  A program of jet veto resummation \cite{Banfi:2012yh, Banfi:2012jm, Berger:2010xi, Tackmann:2012bt, Stewart:2013faa, Becher:2012qa, Becher:2013xia, Moult:2014pja} would in principle be required to isolate these effects. These logs are clearly not taken account in our calculation explicitly due to the fact that there are no jets in our resummation calculation.  Nevertheless, as mentioned earlier, the probability of an event passing the jet veto and the transverse momentum of the \ww system  is strongly correlated, therefore in the process of reweighting the parton showered events and using a jet algorithm, there is a large overlap between the logs accounted for in jet veto resummation, and the logs accounted for in our procedure.  This correlation was observed for instance in~\cite{Banfi:2012yh}, where for Higgs and Drell-Yan the effects of reweighting the \pt distribution agreed very well with the jet-veto efficiency coming from a jet veto resummation calculation.  Given that Higgs production is dominated by gluon initial states, we expect the agreement between reweighting and jet-veto resummation to be even better for \ww. An additional motivation for performing \pt resummation and re-weighting is that we can perform detector simulations on the fully exclusive events, and predict differential observables. It would be interesting to understand the interplay of these effects even further which we leave to future work. 

The rest of the paper is structured as follows.  In Section~\ref{s.resummation}, we outline our methodology and calculation of the NNLL resummed \ww \pt distribution.  In Section~\ref{s.reweighting} we explicitly describe our reweighting procedure, and demonstrate the effects on the total cross section at various energies and compared to various NLO generators and parton showers.  Finally, in Section~\ref{s.discussion}, we discuss the implications of these results both for scale choices used in resummation and the associated errors as well as how to test these effects in other channels.  In particular, given the similarity in scales of \ww, \wz and \zz processes, and the fact that resummation can't tell the difference with respect to the hard matrix element, if resummation effects are responsible for even part of the discrepancy as currently measured there are distinct predictions in other channels.

%%%%%%%%%%%%%%%%%%%%%%%%%%%%%%%%%%%%%%%%%%%%%%%%%%%%%%%%%%%%%%%%%%%%%%
%%%%%%%%%%%%%%%%%%%%%%%%%%%%%%%%%%%%%%%%%%%%%%%%%%%%%%%%%%%%%%%%%%%%%%% 
%%%%%%%%%%%%%%%%%%%%%%%%%%%%%%%%%%%%%%%%%%%%%%%%%%%%%%%%%%%%%%%%%%%%%%
%%%%%%%%%%%%%%%%%%%%%%%%%%%%%%%%%%%%%%%%%%%%%%%%%%%%%%%%%%%%%%%%%%%%%
\section{\ww transverse momentum resummation}
\label{s.resummation} \setcounter{equation}{0} \setcounter{footnote}{0}

\subsection{The resummation method}
For hadron collider production of electroweak bosons with invariant mass $M$ and transverse momentum $p_T$, the fixed-order perturbative expansion acquires powers of large logarithms, $\alpha_s^n \log^m \left( M/p_T \right)$, with $m \leq 2n-1$, which can be resummed to all orders \cite{Dokshitzer:1978yd, Parisi:1979se, Curci:1979bg, Collins:1981uk, Collins:1981va, Kodaira:1981nh, Kodaira:1982az, Altarelli:1984pt, Collins:1984kg, Catani:2000vq}. We implement the method of Refs. \cite{Bozzi:2005wk, Bozzi:2010xn} to calculate the $WW$ transverse momentum distribution at partial NNLL+LO.\footnote{In our convention, LO \pt distribution is at the same $\alpha_s$ order as the NLO total cross section.} Some aspects of the method are outlined below. The factorized cross section is
\begin{align}
\frac {d\sigma^{WW}}{dp_T^2} \left( p_T, M, s \right) &= \sum_{a,b} \int_0^1 dx_1 \int_0^1 dx_2 f_{a/h_1} \left( x_1, \mu_F^2 \right)  f_{b/h_2} \left( x_2, \mu_F^2 \right) \nonumber \\
&\quad \frac{d\hat \sigma^{WW}_{ab}}{dp_T^2} \left(p_T, M, \hat s, \alpha_s \left( \mu_R^2 \right), \mu_R^2, \mu_F^2 \right),\label{eq:factorization}
\end{align}
where $f_{a/h_1}$ and $f_{b/h_2}$ are the parton distribution functions for the parton species $a$ and $b$ in the two colliding hadrons, $\hat s = s x_1 x_2$ is the partonic center of mass energy, and $d \hat \sigma^{WW}_{ab} / dp_T^2$ is the partonic cross section. The partonic cross section will be the sum of a resummed part and a finite part; the finite part matches resummation with fixed order calculations. In our case, we will give partial NNLL+LO results which effectively include the exact LO results at $O\left( \alpha_s \left( \mu_R^2 \right) \right)$, plus partial NNLL resummation correction terms at $O\left( \alpha_s^n \left( \mu_R^2 \right) \right)$, $2 \leq n \leq \infty$. The method of \cite{Bozzi:2005wk, Bozzi:2010xn} ensures that the resummation correction preserves the total cross section (which can be calculated at fixed-order reliably) while improving predictions for differential distributions, especially at low $p_T$.

The quantity that is resummed directly is actually the double transform of the partonic cross section, 
\begin{equation}
\mathcal W_{ab,\,N}^{WW} \left( b, M; \alpha_s \left( \mu_R^2 \right), \mu_R^2, \mu_F^2 \right),
\end{equation}
where $b$, the impact parameter, is the Fourier transform moment with respect to $p_T$, while $N$ is the Mellin transform moment with respect to $z = M / \hat s$.
To invert the Mellin transform, we use the standard formula
\begin{align}
&\quad \mathcal W_{ab}^{WW} \left( b, M, \hat s=  M^2 / z; \alpha_s \left( \mu_R^2 \right), \mu_R^2, \mu_F^2 \right) \nonumber \\
&= \int_{c-i\infty}^{c+i\infty} \frac{dz}{2\pi i}\, z^{-N} \mathcal W_{ab,\,N}^{WW} \left( b, M, \alpha_s \left( \mu_R^2 \right), \mu_R^2, \mu_F^2 \right)
\end{align}
where $c$, a positive number, is the intercept between the integration contour and the real axis. In numerical implementations, the contour is deformed to the left on both the upper and lower complex planes, leaving the integral invariant but improving numerical convergence.  To perform the convolution in Eq. \eqref{eq:factorization}, we fit the parton distribution functions with simple analytic functions \cite{deFlorian:2007sr} to obtain analytical Mellin transforms. We multiply the Mellin transform of the parton distribution functions with the Mellin transform of the partonic cross section, before we actually invert the transform. The error associated with fitting is less than $10^{-3}$.

To invert the Fourier transform in Eq. \eqref{eq:factorization}, we use
\begin{align}
\frac {d \hat \sigma_{ab}^{WW}}{dp_T^2} \left( p_T, M, \hat s, \alpha_s \left( \mu_R^2 \right), \mu_R^2, \mu_F^2 \right) &= \frac {M^2} {\hat s} \int \frac{d^2 \mathbf b} {4\pi} e^{i \mathbf b \cdot p_T} \, \mathcal W_{ab}^{WW} \left( b, M, \hat s, \alpha_s \left( \mu_R^2 \right), \mu_R^2, \mu_F^2  \right) \nonumber \\
&= \frac {M^2} {\hat s} \int \frac{d^2 \mathbf b} {4\pi} \frac b 2 J_0 \left( b p_T \right) \mathcal W_{ab}^{WW} \left( b, M, \hat s, \alpha_s \left( \mu_R^2 \right), \mu_R^2, \mu_F^2  \right).
\label{eq:bessel}
\end{align}
The double transform in Eq. \eqref{eq:factorization} contains large logarithms of the form $\sim \log(M\,b)$ which correspond to $\sim \log \left( M / p_T \right)$ before the Fourier transform. Ignoring the finite term from matching to fixed-order results, the large logarithms are resummed to all order by exponentiation \cite{Bozzi:2005wk},
\begin{align}
\mathcal W_{ab,\,N}^{WW} \left( b, M; \alpha_s \left( \mu_R^2 \right), \mu_R^2, \mu_F^2 \right) &= \mathcal H_N^{WW} \left( M, \alpha_s \left( \mu_R^2 \right); M^2 / \mu_F^2, M^2/Q^2 \right) \nonumber \\
&\quad \times \exp \left\{ \mathcal G_N \left( \alpha_s \left(\mu_R^2\right), L; M^2/\mu_R^2, M^2/Q^2 \right) \right\},
\label{eq:HexpG}
\end{align}
where the $\mathcal H_{NN}^{WW}$ function essentially describes physics at the scale comparable with $M$ and hence does not depend on $b$. To our needed accuracy, the function is deduced from the one-loop QCD virtual correction \cite{Catani:2000vq} for $W^+W^-$ production calculated in \cite{Frixione:1993yp}. On the other hand, the function $\mathcal G_N$ essentially describes physics at the scale of $1/b \sim p_T$ and hence does not depend on the hard process; for example, it is the same for the Drell-Yan process which is also initiated by quark-antiquarks at LO. The quantity $L$ is defined as
\begin{equation}
L \equiv \ln \frac{Q^2 b^2}{b_0^2},\quad b_0 \equiv 2e^{-\gamma_E} \approx 1.12,
\end{equation}
where $Q$, termed the resummation scale, is chosen to be comparable in magnitude to the hard scale of the process. It is an inherent ambiguity in resummation calculations, in addition to the usual $\mu_R$ and $\mu_F$ ambiguities for fixed order calculations.

The exponent in Eq. \eqref{eq:HexpG} can be expanded in successive logarithmic orders \cite{Bozzi:2005wk,Catani:2003zt}
\begin{align}
\mathcal G_N \left( \alpha_s, L; M^2/\mu_R^2, M^2/Q^2 \right) &= L g^{(1)} \left( \alpha_s L \right) + g_N^{(2)} \left( \alpha_s L; M^2 / \mu_R^2, M^2/Q^2 \right) \nonumber \\
& \quad + \frac{\alpha_s}{\pi} g_N^{(3)} \left( \alpha_s L; M^2 / \mu_R^2, M^2/Q^2 \right)  \nonumber \\
&\quad + \sum_{n=4}^{+\infty} \left( \frac{\alpha_s}{\pi} \right)^{n-2} g_N^{(n)} \left( \alpha_s L; M^2 / \mu_R^2, M^2/Q^2 \right).
\label{eq:LL-NLL-NNLL}
\end{align}
This expansion makes sense if we regard $\alpha_s L$ as of order unity. The $g^{(1)}$ term is the leading logarithmic (LL) term, while $g^{(2)}$ and $g^{(3)}$ are the NLL and NNLL terms, and so on. 
The variation of Q shuffles terms between the fixed order and resummed terms and can give an estimate for as yet uncomputed higher Logs.
%To actually compute the coefficients in Eq. \eqref{eq:LL-NLL-NNLL}, we make use of the following formula \cite{Collins:1984kg Catani:2000vq},
%
%\begin{align}
%\mathcal W_N^{WW} &= \sum_{c} \sigma_{c\bar c,\, WW}^{(0)} \left( \alpha_s \left( M^2 \right), M \right) H_c^{WW} \left( \alpha_s \left( M^2 \right) \right) S_c \left( M,b \right)
%\end{align}
%

The necessary ingredients to perform NLL resummation can be found in \cite{Bozzi:2005wk, Catani:2000vq}. In addition, we also include the three-loop coefficients $A^{(3)}$ for the Sudakov form factor, calculated in \cite{Becher:2010tm}, to achieve approximate NNLL accuracy. We re-used part of the QCD-Pegasus code \cite{Vogt:2004ns} to calculate the NLO splitting kernel in complex moment space.
\subsection{Numerical results}
\begin{figure}[ht!]
\begin{centering}
\includegraphics[scale=0.3]{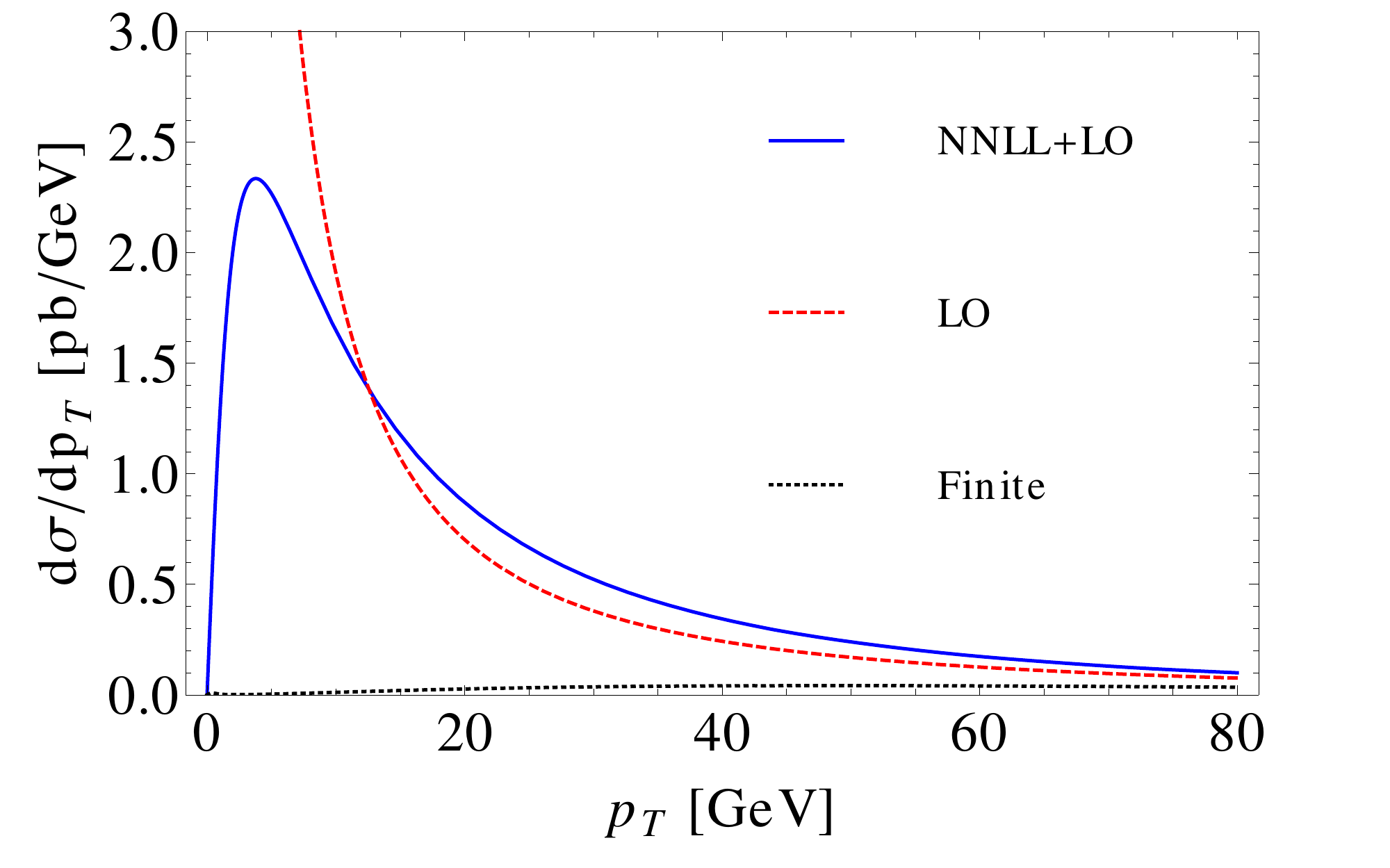}
\par\end{centering}
\caption{Plot of resummed, finite (matching) and fixed-order $W^+W^-$ transverse momentum distributions from 8 TeV proton collisions.}
\label{fig:8TeV-resum-fixed-finite}
\end{figure}
The full details about the underlying resummation formalism, including the diagonalization of the DGLAP splitting kernel in the multi-flavor case, and the matching to fixed-order calculations, are covered in  \cite{Bozzi:2005wk, Bozzi:2010xn} and will not be repeated here. We now go on to present numerical results. To make sure our numerical implementation is correct, we have reproduced the Z-boson resummed transverse momentum distribution in \cite{Bozzi:2010xn}, including effects of varying the resummation scale $Q$.

\begin{figure}[h]
\begin{centering}
\includegraphics[scale=0.4]{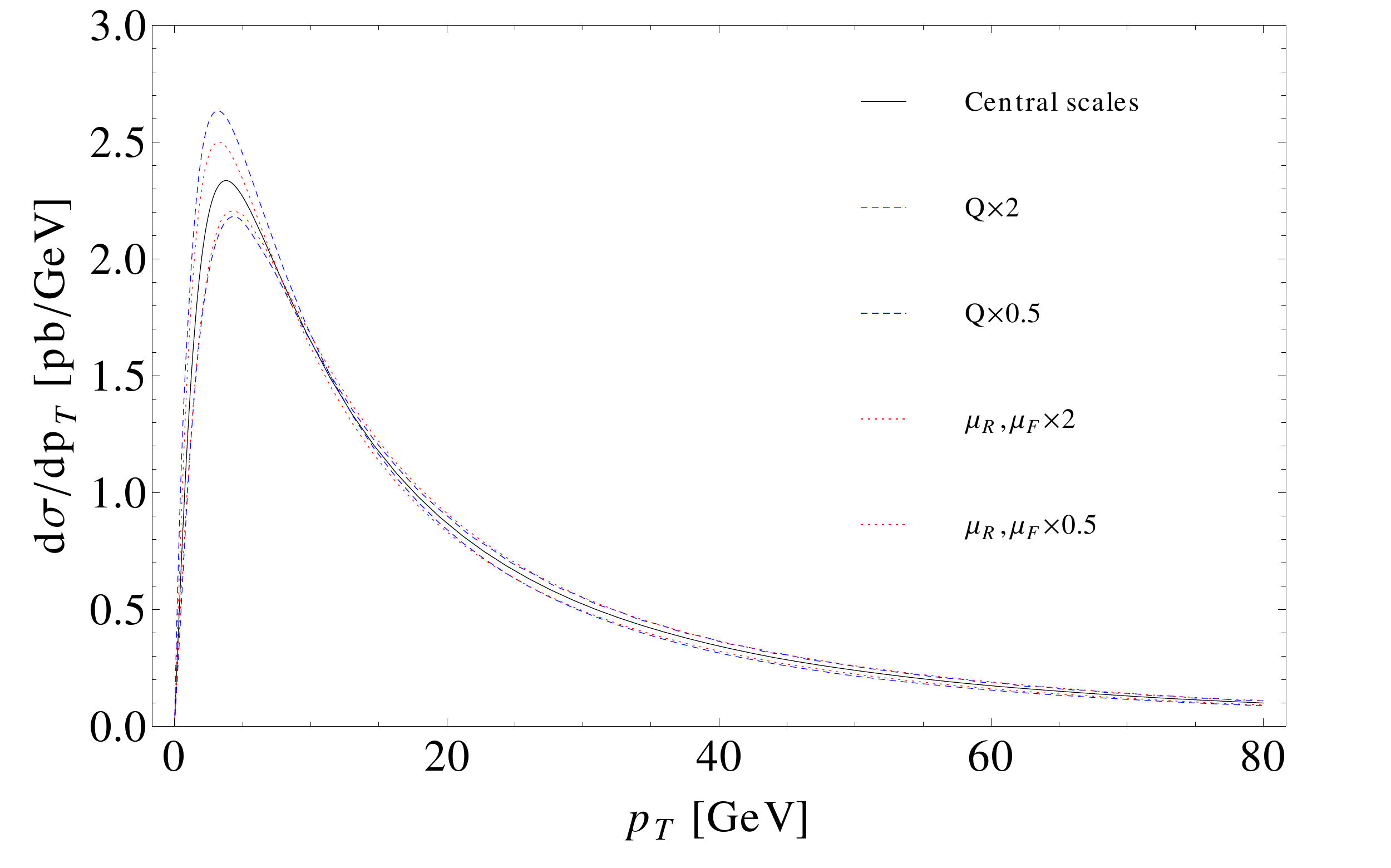}
\par\end{centering}
\caption{Plot of renormalization, factorization and resummation scale variations of the $W^+W^-$ transverse momentum distribution for 8 TeV collisions.}
\label{fig:8TeV-vary}
\end{figure}

We use the MSTW 2008 NLO parton distribution functions \cite{Martin:2009iq}. The central scales we use are $\mu_R=\mu_F = 2m_W$, $Q=m_W$. In fig. \ref{fig:8TeV-resum-fixed-finite}, we plot the resummed, fixed-order, and finite part of the $W^+W^-$ transverse momentum distribution using central scales for 8 TeV pp collisions. We can see that resummation cures the $p_T \to 0$ divergence of the LO distribution and generates substantial corrections. The total cross section obtained from integrating our $p_T$ distribution agrees with exact fixed order results to better than $0.5\%$, which is a consistency check for our numerical accuracy.

To assess perturbative scale uncertainties, we simultaneously vary $\mu_R$ and $\mu_F$ up and down by a factor of 2, and separately vary $Q$ up and down by a factor of 2. The resulting variations in the transverse momentum distributions are plotted in Fig. \ref{fig:8TeV-vary} for 8 TeV collisions. We can see that the largest scale uncertainties result from varying the resummation scale $Q$.
\begin{figure}
\begin{centering}
\includegraphics[scale=0.25]{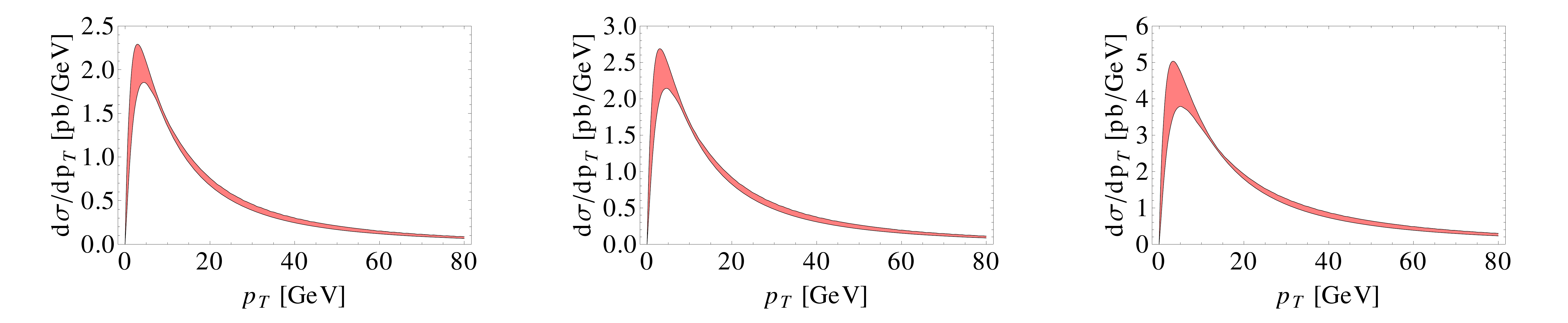}
\par\end{centering}
\caption{NNLO+LO predictions, with error bands, for the $W^+W^-$ transverse momentum distribution for 7,8 and 14 TeV collisions.}
\label{fig:7-8-14TeV-errorband}
\end{figure}
By adding $\mu_R$ \& $\mu_F$ variations and $Q$ variations in quadrature, we produce the distribution with error bands, for 7, 8 and 14 TeV, shown in Fig. \ref{fig:7-8-14TeV-errorband}. The combined scale uncertainty at the peak of the distribution is around $\pm 10\%$ for each collision energy.
\begin{figure}[h]
\begin{centering}
\includegraphics[scale=0.3]{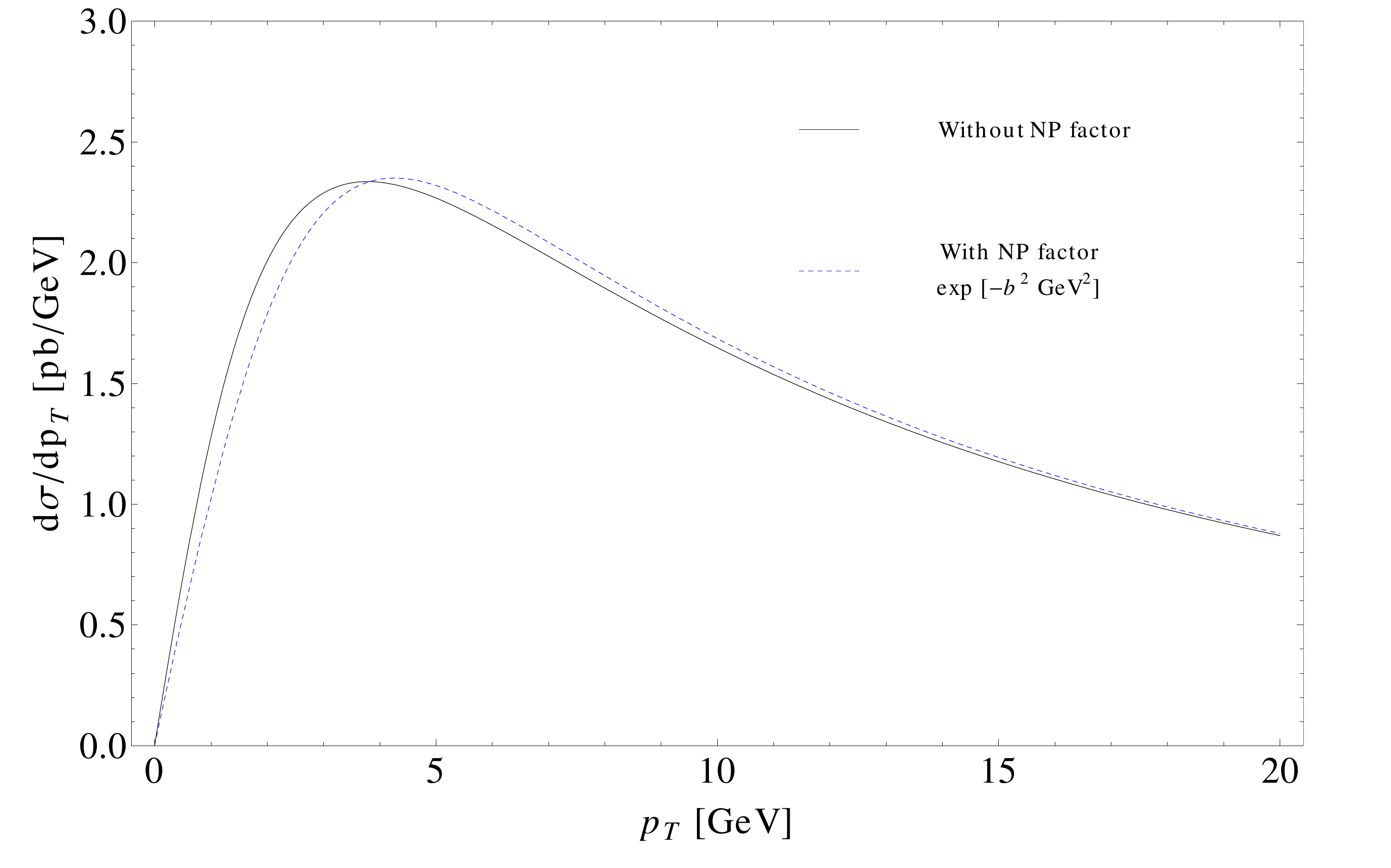}
\par\end{centering}
\caption{NNLL+LO prediction for the WW transverse momentum distribution at 8 TeV, with and without the non-perturbative Gaussian smearing factor $\exp \left[ -1\, {\rm GeV}^2\, b^2 \right]$.}
\label{fig:NP}
\end{figure}

We now briefly mention non-perturbative effects. In Eq. \eqref{eq:bessel} $\mathcal W_{ab}^{WW}$ in fact becomes singular at large $b$ due to the divergence of the QCD running coupling below the scale $\Lambda_{\rm QCD}$. This is a non-perturbative issue and becomes important at low $p_T$. Many prescriptions for regulating the non-perturbative singularity exists, such as the $b^*$ model \cite{Collins:1981va,Collins:1984kg} and the minimal prescription \cite{Laenen:2000de}. We adopt a simple cutoff at $b=2\gev^{-1}$, and give results both with and without an additional non-perturbative Gaussian smearing factor of $\exp \left[ -g_{NP}^2\, {\rm GeV}^2\, b^2 \right]$ with $g_{NP} =1$. The $W^+W^-$ fiducial cross sections after reweighting parton shower events shifts only differ by about 1\% with and without the Gaussian smearing factor, much smaller than the perturbative scale uncertainties we will encounter. In Fig. \ref{fig:NP} we compare the predicted WW transverse momentum distribution with and without the Gaussian smearing factor. The smearing causes the peak to shift by about 0.5 GeV to larger $p_T$.

\begin{figure}[h]
\begin{centering}
\includegraphics[scale=0.55]{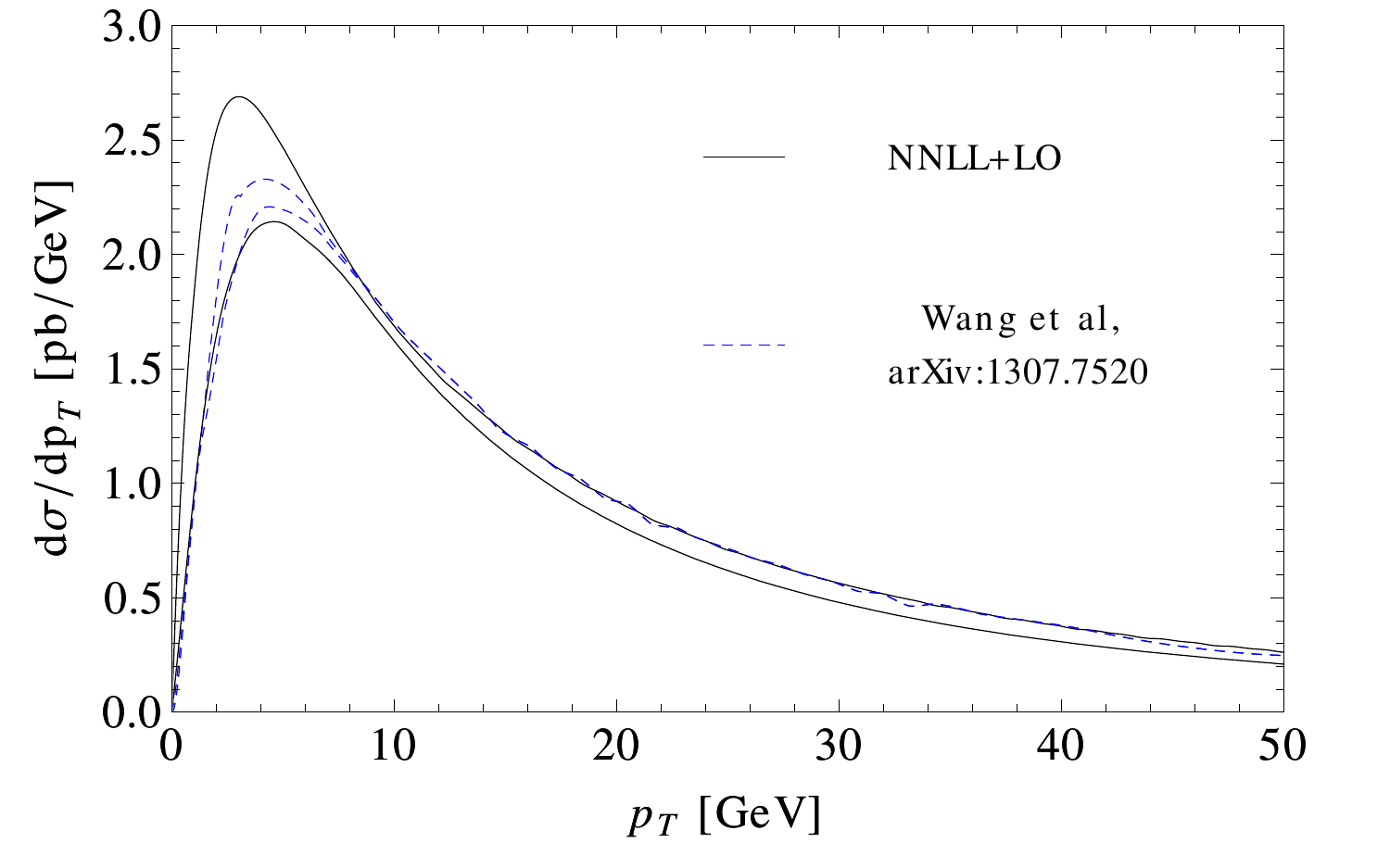}
\par\end{centering}
\caption{Comparision of our resummed WW $p_T$ distribution with a SCET-based resummation calculation, with error bands shown for both.}
\label{fig:scet}
\end{figure}

Finally, we compare our $p_T$ distribution at 8 TeV with the SCET-based resummation calculation by \cite{Wang:2013qua} in Fig. \ref{fig:scet}. The results are in good agreement, but our results show a larger error band because we varied both $\mu_R$ (with $\mu_F$ locked to be equal to $\mu_R$) and the resummation scale $Q$, the latter of which indicates ambiguities in splitting contributions into the resummed part and the finite part, while the calculation by \cite{Wang:2013qua} only considers the variation of one scale.

%%%%%%%%%%%%%%%%%%%%%%%%%%%%%%%%%%%%%%%%%%%%%%%%%%%%%%%%%%%%%%%%%%%%%%
%%%%%%%%%%%%%%%%%%%%%%%%%%%%%%%%%%%%%%%%%%%%%%%%%%%%%%%%%%%%%%%%%%%%%%% 
\section{Transverse Momentum Reweighting and Fiducial Cross Sections}
\label{s.reweighting} \setcounter{equation}{0} \setcounter{footnote}{0}
%%%%%%%%%%%%%%%%%%%%%%%%%%%%%%%%%%%%%%%%%%%%%%%%%%%%%%%%%%%%%%%%%%%%%%
%%%%%%%%%%%%%%%%%%%%%%%%%%%%%%%%%%%%%%%%%%%%%%%%%%%%%%%%%%%%%%%%%%%%%

The transverse momentum resummation shown in Section~\ref{s.resummation} systematically improves our understanding of the \pt distribution of the diboson system.  However, the \ww \pt distribution as measured by the LHC experiments is not the same as the distribution that is calculated in Section~\ref{s.resummation}. This is because the detector only measures a certain fiducial region of phase space, there are additional cuts put on the physics objects to reduce backgrounds, and finally there are detector effects which smear the \pt distribution compared to the theoretical prediction.  In very clean channels such as Drell-Yan or \zz production, these effects can be unfolded more easily, and an unambiguous prediction for the \pt of the system can be compared to theoretical predictions.  For \ww the effects are more difficult to unfold and as of yet a full analysis has yet to be compared to the experimental results for the \ww diboson system's \pt.  In fact, only ATLAS has released a distribution, the vector sum of the \pt of the leptons and MET, directly correlated to the \pt of the diboson system.

In order to compare to data, we must implement the same cuts that the experiments perform.  Immediately this runs into potential problems as the distributions predicted in Section~\ref{s.resummation} are fully inclusive, and even at the leptonic level there are cuts that restrict the dsitributions to a fiducial phase space.  To circumvent these difficulties we implement a reweighting procedure on generated events for the \pt of the system prior to cuts, and then perform the analysis cuts to find the effects of \pt resummation.   This of course isn't a perfect matching of the effects of resummation and data, but without unfolded distributions this is the closest possible comparison that can be made at this point.   This procedure is akin to that used for predicting the Higgs signal at the LHC, where the transverse momentum resummed shape, taken from HqT for instance\cite{hqt}, is used to reweight the MC simulated events.

\begin{figure}[ht]
\begin{centering}
\includegraphics[scale=0.75]{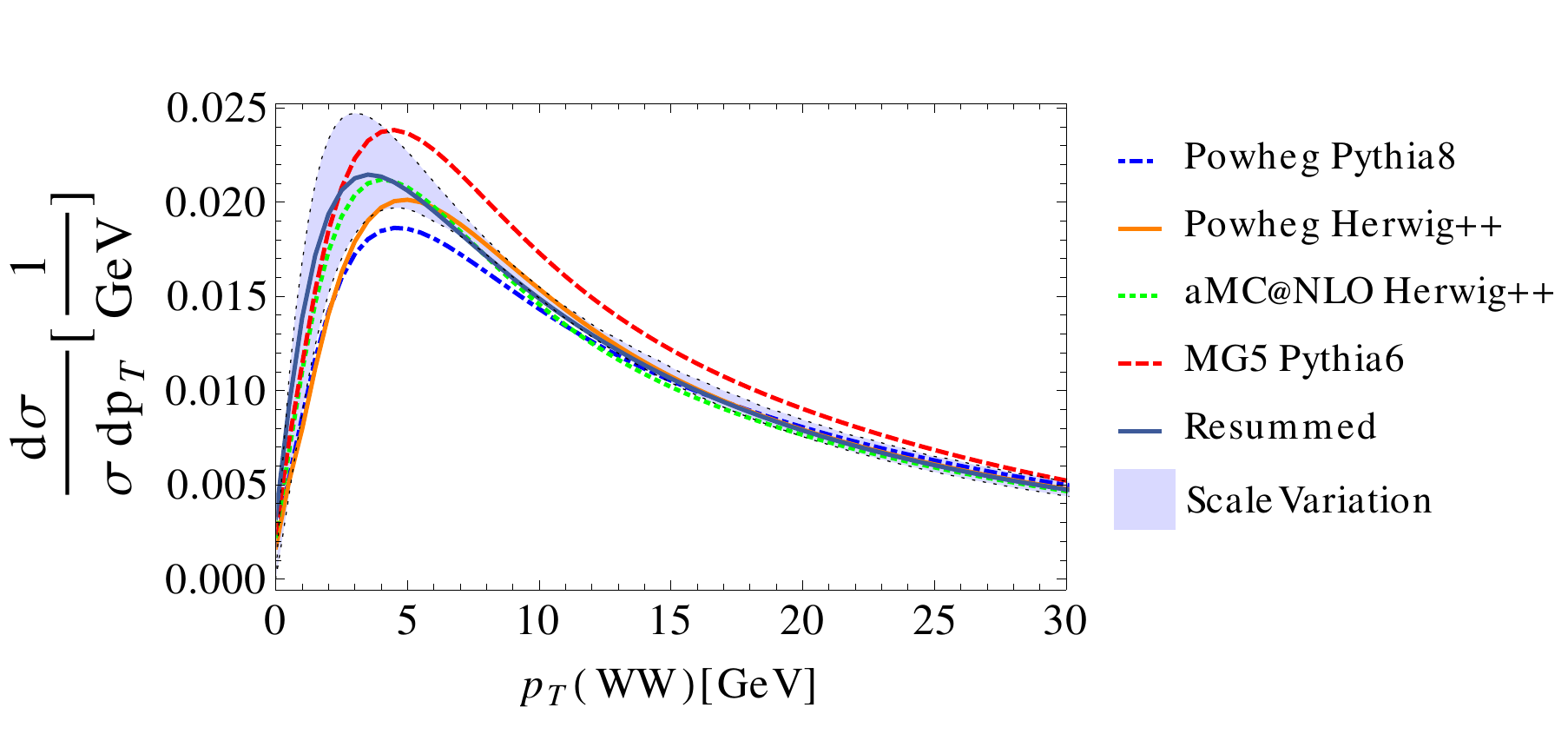}
\par\end{centering}
\caption{Plot of Resummation predicted and MC+shower predictions for $W^+W^-$ transverse momentum distributions at 8 TeV.  The shaded region represents the scale $Q$ variation by a factor of 2 relative to the central scale choice $Q=m_W$ for the resummation prediction.}
\label{fig:8TeVMCshapes}
\end{figure}

It is possible that a comparison between reweighted events after experimental cuts and the original Monte Carlo events {\em could} predict the same cross section.  The formalism we use by definition does not change the total inclusive cross section.  However, if the reweighted distributions that have a different shape are also cut on, then this will effect the total measured cross section. This happens because the cuts  change the fiducial cross section and hence the inferred total cross section once the acceptances and efficiencies are unfolded.  As we will show, there isn't a direct cut on the reweighted \pt distribution, but the jet veto cut is highly correlated with it and significantly effects the extrapolated total cross section.  Additionally, the cause of the correlation will also reflect that different underlying Monte Carlo generators and parton showers will have different size effects when extrapolating to the total cross section.  These differences are demonstrated in Figure~\ref{fig:8TeVMCshapes} where the \pt distributions predicted by resummation are compared to various Monte Carlos (POWHEG BOX\cite{Nason:2004rx, Frixione:2007vw, Alioli:2010xd}, MadGraph/aMC@NLO and matched Madgraph 0j+1j\cite{Alwall:2014hca}) in combination with  different parton showers from Herwig++\cite{Bahr:2008pv} and Pythia8\cite{Sjostrand:2006za}. MSTW2008 NLO pdf sets were used for all NLO event generations to be consistent with resummaton and CTEQ6 LO pdf\cite{Pumplin:2002vw} was used for the Madgraph 0+1j analysis. 

To perform the reweighting procedure, resummed theory curves from Section~\ref{s.resummation} and MC events are binned into 0.5 GeV bins along $p_{T_{WW}}$. A reweighting factor 
is then computed
\begin{equation}
F[p_T]=\frac{\textrm{Resummed\,bin}[p_T]}{\textrm{MC\,bin}[p_T]}.
\label{eq:reweighting factor}
\end{equation}
To approximate detector effects MC events are then smeared using Delphes\cite{deFavereau:2013fsa} for a fast detector simulation\footnote{The detector simulation is important to match data, as the \pt distribution of the diboson system predicted by MC@NLO\cite{Frixione:2002ik} shown by ATLAS can not be matched without additional smearing of the MET.  We demonstrated this with both PGS and Delphes.  In the end however, this smearing does not effect the resummation reweighting effects shown here, because the underlying MC events and resummed reweighted events are affected in the same way.  We have demonstrated this explicitly by changing the MET resolution by a factor of 2 each way, which simply shifts the peak of the \pt distribution.}.  Finally, once detector level events are produced we apply the cuts performed by the LHC experiments.  An example of the cuts implemented by the ATLAS measurement at 7 TeV  is reproduced below in Table~\ref{table:cuts}.  
The cuts from CMS are quite similar, the jet veto as we will show turns out to be the most important effect, and CMS has a jet veto of $30$ GeV compared to $25$ GeV for ATLAS.  We comment on this slight difference in Section~\ref{s.discussion}, however, since CMS hasn't produced a plot of the \pt of the \ww system similar to ATLAS, we adopt the ATLAS cuts when demonstrating the effects of using the \pt resummed reweighted distributions. Pythia8 was used with default tuning and since all our results are shape dependent, the reweighting procedure should be performed again using our resummation-theory curves when using a non-default pythia8 tuning.

\begin{table}[htb]
\begin{centering}
  \begin{tabular}{|c|}
    \hline
    Exactly two oppositely-sign leptons, $p_T>20\gev, {p_T}_{\rm leading}>25\gev$ \\ \hline
      $m_{ll'} >15,15,10\gev $ (ee,$\mu\mu$,$e\mu$)  \\ \hline
      $\lvert m_{ll'}-m_Z \rvert > 15,15,0\gev$ (ee,$\mu\mu$,$e\mu$) \\ \hline
      $E_{T,\rm{Rel}}^{\rm miss}>45,45,25\gev $(ee,$\mu\mu$,$e\mu$) \\ \hline
     Jet Veto $25\gev$\\ \hline
    ${p_T}_{ll'}>30\gev$  \\ \hline

\end{tabular}
\caption[Table caption text]{ATLAS cut flow for 7 TeV analysis\cite{atlas7ww}}
\label{table:cuts}
\end{centering}
\end{table}

\subsection{Reweighting Results}
 
 We perform the reweighting as described above using a central scale $Q=m_W$ as well as varying the resummation scale Q up and down by a factor of 2 while keeping $\mu_R$ and $\mu_F$ fixed.  We define the percentage difference caused by reweighting as 
 \begin{equation}
 \text{percentage difference}=\frac{(\text{events}_{\rm res}-\text{events}_{\rm MC})\cdot 100}{\text{events}_{\rm MC}}
\end{equation}
where 
\begin{itemize}
\item $\text{events}_{MC}$ is events predicted by the MC before reweighting
\item $\text{events}_{res}$ is events after reweighting the MC events.
\end{itemize}
with a positive percentage difference implying an increase in the theoretical prediction on $\sigma_{\text{Fid}}$. It is important to notice that reweighting is done with respect to $p_{T_{WW}}$ just after the shower but before detector simulation.  To demonstrate the effects of other scale variations on  $\sigma_{\text{Fid}}$ we also varied  $\mu_R$ and $\mu_F$ as well as 
the non-perturbative factor discussed in Section~\ref{s.resummation} and report the percentage differences compared to Powheg +Pythia8 (8 TeV) as an example in Table~\ref{table:PP8}.
\begin{center}
\begin{table}[htb]
\begin{centering}
  \begin{tabular}{|c|c|c| }
    \hline
   Scale Choice & \% difference & \% difference with $g_{\rm NP}=1$ \\ \hline
   Combined &${6.5}^{+5.0}_{-3.0}$  &${6.4}^{+5.0}_{-3.0}$ \\ \hline
   Central scales, $Q=m_W$, $\mu_R=\mu_F=2m_W$ & 6.51  & 6.38 \\ \hline
   $Q=2\times$central  & 4.96 & 4.82 \\ \hline
   $Q=0.5\times$central  & 10.75 & 10.64 \\ \hline
   $\mu_{R}=\mu_{F}=0.5\times$central  & 3.89 &3.76  \\ \hline
   $\mu_{R}=\mu_{F}=2\times$central  & 9.16 & 9.04 \\ \hline
\end{tabular}
\caption{Percentage differences of reweighted theory predictions compared to Powheg+Pythia8 at 8 TeV for $\sigma_{\text{Fid}}$  and various choices of scale.  The 2nd column does not include the Gaussian smearing factor for non-perturbative effects, while the 3rd column includes a non-zero non-perturbative factor $g_{NP}=1$ typical for quark dominated initial states.}
\label{table:PP8}
\end{centering}
\end{table}
\end{center}
We find that as observed in Section~\ref{s.resummation}, the Q variation leads to a larger percentage difference than the $\mu_F$ or $\mu_R$ scale variation.  The non-perturbative factor $g_{NP}$ shifts the peak of the underlying \pt distributions slightly, but in the end has a minimal effect on the cross section.  We show the effects of reweighting on MC generators and parton showers in Tables ~\ref{table:tab7}, ~\ref{table:tab8}, ~\ref{table:tab14} for 7,8 and 14 TeV respectively.

\begin{center}
\begin{table}[ht]
\begin{centering}
  \begin{tabular}{|l||c| }
    \hline
    MC + Parton Shower & Corrections (\%) \\ \hline
    Powheg+Pythia8   & ${6.4}^{+4.7}_{-2.8}$ \\ \hline
     Powheg+Herwig++ & ${3.8}^{+4.5}_{-2.6}$  \\ \hline
    aMC@NLO+Herwig++ &  ${3.3}^{+5.0}_{-3.0}$ \\ \hline
\end{tabular}
\caption{Percentage differences for $\sigma_{\text{Fid}}$ of reweighted theory predictions compared to MCs+Parton Showers at 7 TeV.}
\label{table:tab7}
\end{centering}
\end{table}
\end{center}

\begin{center}
\begin{table}[ht]
\begin{centering}
  \begin{tabular}{|l||c| }
    \hline
    MC + Parton Shower & Corrections (\%) \\ \hline
    Powheg+Pythia8   & ${6.5}^{+5.0}_{-3.0}$ \\ \hline
    Powheg+Herwig++ & ${3.8}^{+4.3}_{-2.5}$  \\ \hline
    aMC@NLO+Herwig++ &  ${3.1}^{+5.0}_{-3.0}$ \\ \hline
    MADGRAPH LO+Pythia6 & ${-9.6}^{+4.4}_{-2.7}$ \\ \hline
\end{tabular}
\caption{Percentage differences for $\sigma_{\text{Fid}}$ of reweighted theory predictions compared to MCs+Parton Showers at 8 TeV.}
\label{table:tab8}
\end{centering}
\end{table}
\end{center}

\begin{center}
\begin{table}[ht]
\begin{centering}
  \begin{tabular}{|l||c| }
    \hline
    MC + Parton Shower & Corrections (\%) \\ \hline
    Powheg+Pythia8   & ${7.0}^{+6.4}_{-5.1}$ \\ \hline
     Powheg+Herwig++ & ${4.4}^{+5.9}_{-4.7}$  \\ \hline
    aMC@NLO+Herwig++ &  ${4.2}^{+6.5}_{-5.2}$ \\ \hline
\end{tabular}
\caption{Percentage differences for $\sigma_{\text{Fid}}$ of reweighted theory predictions compared to MCs+Parton Showers at 14 TeV.}
\label{table:tab14}
\end{centering}
\end{table}
\end{center}

\begin{figure}[ht!]
\begin{centering}
\includegraphics[scale=0.35]{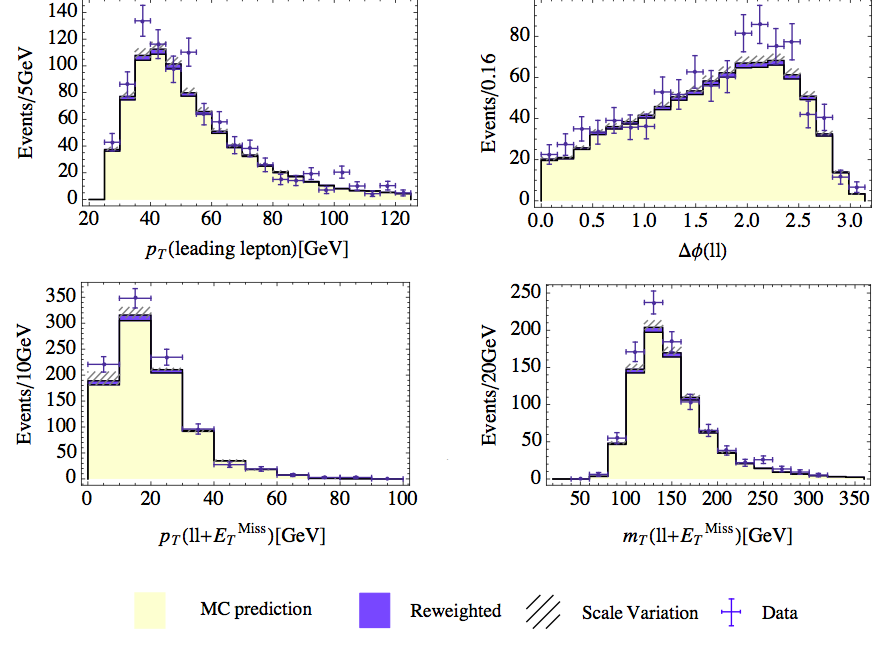}
\par\end{centering}
\caption{ aMC@NLO+Herwig++ observables histogrammed for $W^+W^-$ transverse momentum distribution for 7 TeV collisions and including the reweighting correction. }
\label{fig:7tevreweight}
\end{figure}

To demonstrate the effects on differential distributions, we use the ATLAS cutflows and show the predictions of \pt resummation for the 7 TeV ATLAS study\cite{atlas7ww} compared to the original MC@NLO+Herwig++ results used by ATLAS. In Figure~\ref{fig:7tevreweight}, we plot the four distributions shown in~\cite{atlas7ww}.  As can be seen in Figure~\ref{fig:7tevreweight}, \pt reweighting can improve the differential distributions somewhat, but is not capable of explaining the full discrepancy using a central choice of scales. 

\begin{figure}[htb]
\begin{centering}
\includegraphics[scale=0.4]{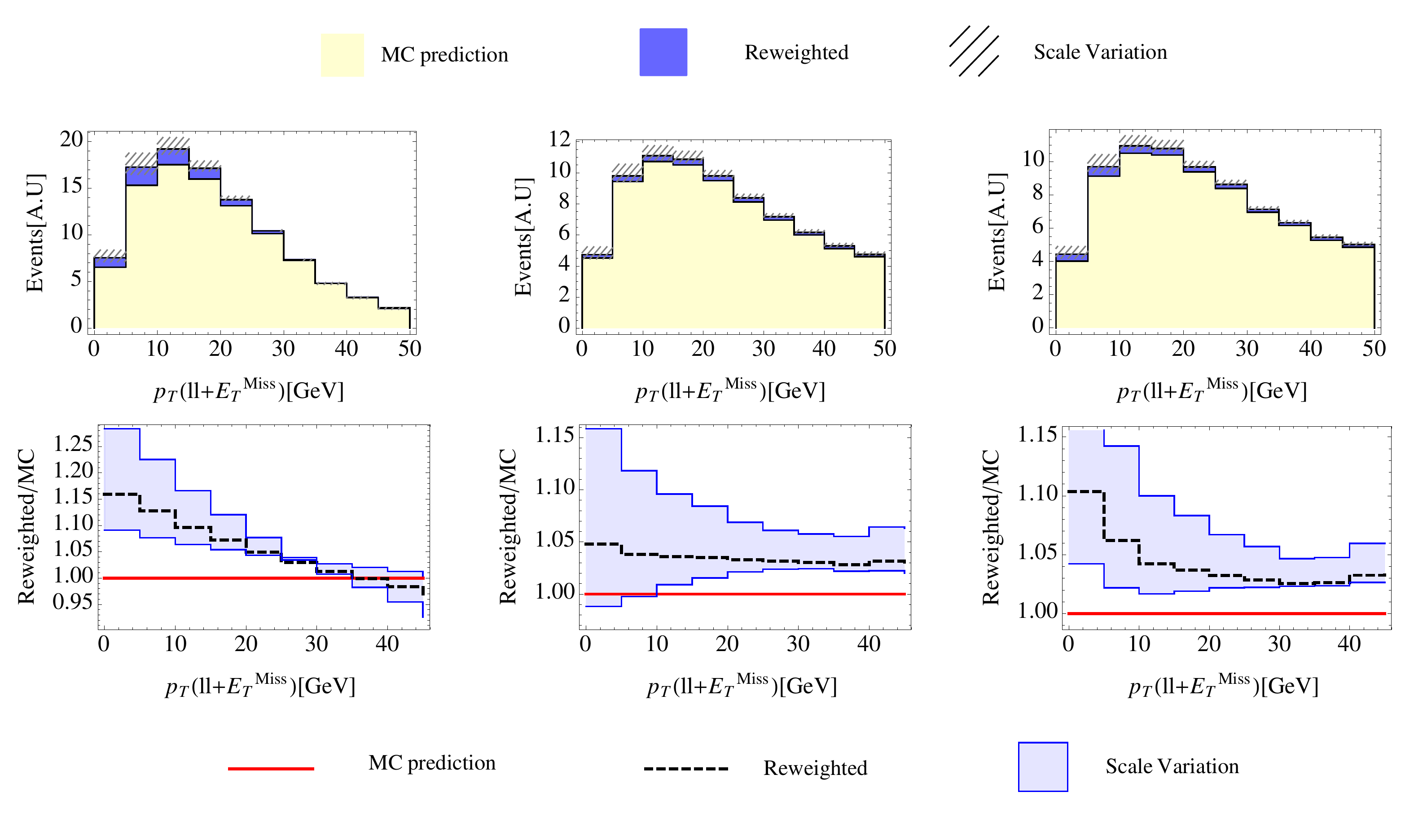}
\par\end{centering}
\caption{The top row shows the reweighting correction for left (Powheg+Pythia8), center (aMC@NLO+Herwig++), right (Powheg+Herwig++) to the $p_{T}(ll+E_T^{\rm miss})$ observable. The bottom row has bin-by-bin percentage difference in events between reweighting and the MC + PS. }
\label{fig:8PPAHPH}
\end{figure}

 To demonstrate the effects at 8 TeV we show the distribution most affected, $p_T(ll+E_T^{\rm miss})$, in Figure~\ref{fig:8PPAHPH} using the same cutflows and different generators.  This distribution is directly correlated with the \pt of the diboson system predicted by resummation, and shows the variation compared to MC generators + parton showers.  The largest discrepancy compared to MC comes from the use of Powheg+Pythia8, while both Powheg and aMC@NLO are in much better agreement when Herwig++ is used as the parton shower.  However, this does not mean the effects of the parton shower are the sole cause of the discrepancy.  In the fractional difference shown in Figure ~\ref{fig:8PPAHPH}, we see the roughly the same shape dependence for both Powheg curves, but the overall magnitude is reduced for Powheg+Herwig++ compared to Powheg+Pythia8.

\subsection{Jet Veto}
\begin{center}   
\begin{figure}[htb]
\begin{centering}
\includegraphics[scale=0.4]{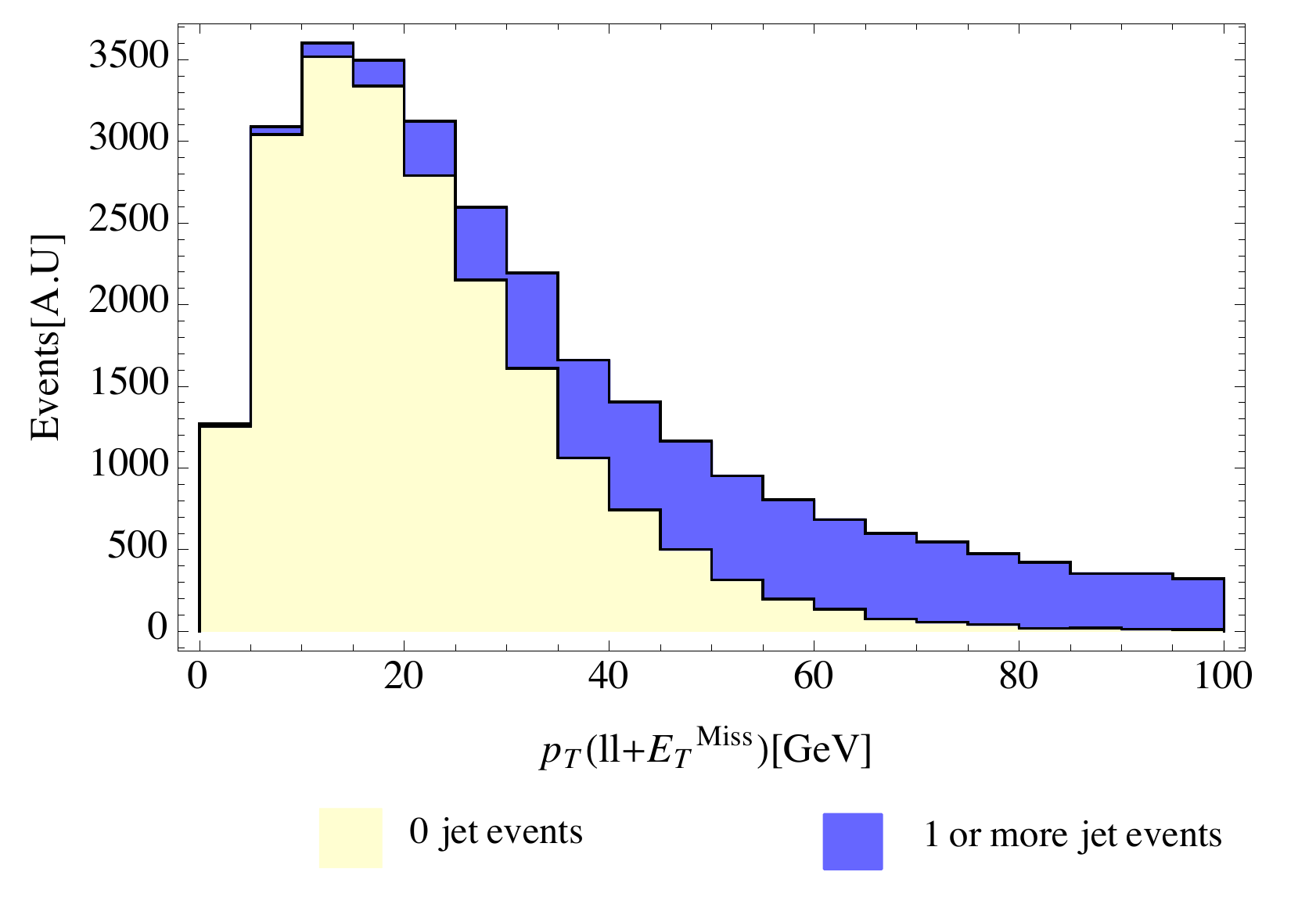}
\par\end{centering}
\caption{Events before the Jet veto. The number of 0 jet events or events with 1 or more jets is shown as a function of the \pt of the diboson system.  Since 1 or more jet - events are vetoed, this sculpts the \pt-shape.}
\label{fig:JetVeto}
\end{figure}
\end{center}

As we have shown thus far, even though the inclusive total cross-sections are the same by design, there are appreciable corrections to the fiducial cross section after reweighting. This means that some of the cuts are well correlated with the ${p_T}_{WW}$ variable and seem to preferentially select the low ${p_T}_{WW}$ region where the resummation curve dominates all the MCs except Madgraph LO. The percentage change due to reweighting at each cut level was analyzed, and as an example the effects of reweighting at each state in the cut flow is shown for Powheg-Pythia8 at 8 TeV in Table \ref{table:cutweight}.  The jet veto stage is the largest contributor to the reweighting excess. To explicitly check this, the order of the jet veto and $p_{Tll}$ cuts was reversed and the biggest jump was found to still be the jet veto cut.   In Figure \ref{fig:JetVeto} we show the correlation between 0 jet events and $>0$ jet events as a function of $p_T(ll+E_T^{\rm miss})$  before the jet veto is applied.  Note that in Figure~\ref{fig:JetVeto}, 0-jet events primarily comprise the low \pt of the diboson system, and as such a jet veto implies that the fiducial cross section will become more sensitive to the shape given by \pt resummation. This clearly points to the Jet Veto cut as the major contributor to changes in the fiducial cross section from \pt resummation reweighting.  If the jet veto were increased this result would still hold, however the 0-jet cross section would then be integrated over a larger range of \pt for the diboson, and thus there would be a smaller effect on the fiducial cross section.  In particular, if the jet veto were dropped entirely this would be equivalent to integrating over the entire diboson \pt which by definition would not change the measured cross section.

\begin{center}
\begin{table}[htb]

\begin{centering}
  \begin{tabular}{|c|c|}
    \hline
    {\bf Cut} & {\bf \% difference} \\ \hline
    Exactly two oppositely-sign leptons, $p_T>20\gev, {p_T}_{\text{leading}}>25\gev$ & 1.36 \\ \hline
      $m_{ll'}$ cuts & 1.16  \\ \hline
      $E_{T,\rm{Rel}}^{\rm miss}$ & 0.83 \\ \hline
     Jet Veto & 9.72 \\ \hline
    ${p_T}_{ll'}$  & 10.75 \\ \hline
\end{tabular}
 \caption{Percentage increase due to Resummation-Reweighting ($Q=\frac{m_W}{2}$, $\mu_R=\mu_F=2m_W$) compared to Powheg-Pythia8 at 8 TeV for each cut stage in the cutflows listed from Table~\ref{table:cuts}.  All percentages are cumulative showing that the jet-veto is the largest effect.}
\label{table:cutweight}
\end{centering}
\end{table}
\end{center}
    
%%%%%%%%%%%%%%%%%%%%%%%%%%%%%%%%%%%%%%%%%%%%%%%%%%%%%%%%%%%%%%%%%%%%%%
%%%%%%%%%%%%%%%%%%%%%%%%%%%%%%%%%%%%%%%%%%%%%%%%%%%%%%%%%%%%%%%%%%%%%%% 
\section{Discussion}
\label{s.discussion} \setcounter{equation}{0} \setcounter{footnote}{0}
%%%%%%%%%%%%%%%%%%%%%%%%%%%%%%%%%%%%%%%%%%%%%%%%%%%%%%%%%%%%%%%%%%%%%%
%%%%%%%%%%%%%%%%%%%%%%%%%%%%%%%%%%%%%%%%%%%%%%%%%%%%%%%%%%%%%%%%%%%%%

As we have shown, \pt resummation, when used to reweight NLO MC distributions, can have a sizable effect on the predicted fiducial and the inferred total cross sections. The general trend in comparison with Monte Carlo generators and parton showers is to increase the predicted cross section $\sim 3-7\%$ and thus decrease the observed discrepancy compared to ATLAS and CMS.  However, this statement depends on the choice of resummation scale for the \ww final state.  At large \pt the fixed order calculation is valid , while at small to moderate \pt the resummation calculation is most reliable.  This scale in practice is analogous to the Matrix Element-Parton Shower matching scale when implementing matching procedures between the two.  As discussed in Section~\ref{s.resummation}, the resummation scale should be similar to the other hard scales in the problem.  We have chosen the simple scale choice, analogous to what is done for Drell-Yan\cite{Bozzi:2008bb,Bozzi:2010xn}, of $\sim M/2$, which for the \ww process we have approximated as the fixed scale $Q=M_W$.  We have demonstrated that the variation in this scale actually can imply quite a deal of uncertainty.  For instance at 8 TeV using  Powheg+Pythia8, by varying $Q$ by a factor of 2 each we introduce a variation on the measured cross section~$\sim \pm 3\%$.  

Given that there is no a priori correct choice of the scale $Q$, a question naturally arises whether one can simply choose a scale to match the experimental data presented thus far.  It is important to note that the measurements presented thus far have used {\em different} event generators  and parton showers.  For instance the preliminary result at 8 TeV by CMS\cite{cms8ww} used Madraph LO + Pythia6 whereas the ATLAS full luminosity result~\cite{atlas8ww} used Powheg + Pythia8.  As shown in Table \ref{table:tab8}, the excess shown by CMS should be even {\em larger} based on our results while the ATLAS discrepancy should be reduced as mentioned above.  Therefore even if a single scale was chosen for the results thus far, it wouldn't imply that it could put both experiments results into better agreement with the SM.  However, if a consistent choice of generator was implemented we could in principle address the question of choosing a scale that was best for this process.    

If one chooses a best fit scale $Q$ to fit the experimental discrepancy in \ww data there are potential implications elsewhere.  Given that the premise behind resummation is that it should be approximately factorized from the hard process, if the initial partons for two processes are the same and there are no colored particles that are exclusively identified in the final state, the effects of resummation should be universal for different processes with similar scales.  Thus, if there is a ``correct" choice of scale for the \ww process, then this choice of scale should be implemented for \wz and \zz diboson process as well, because of the similar hard scales in these diboson systems.  The \wz and \zz process are experimentally even easier channels, especially the \zz channel where the  \pt of the \zz system can be reconstructed with less uncertainty.  The drawback of course is the reduced number of events in these channels, but nevertheless statistics are starting to approach the point where a useful comparison can be made e.g. the recent CMS 8 TeV measurement~\cite{cms8ww}.   The \zz cross section and \pt distribution \cite{Khachatryan:2014dia} are in remarkable agreement with the SM and as such if there were a large change in the \pt distribution caused by the use of the best-fit \ww scale $Q$ then it would cause the agreement with the SM of \zz to become worse. However, we note that the change from \pt resummation in the \ww process is mostly due to the imposition of a jet-veto which the \zz channel does not have. We find that choosing a scale that fits the \ww discrepancy and naively calculating the inclusive change for \zz causes a disagreement with data in all \pt bins.  Further study of course is warranted, and a simultaneous fit should be employed to understand the agreement between the SM and measured diboson processes.  This of course brings up a more general point that in analyzing the agreement between the SM and LHC data, similar theoretical methods should be employed and not just a choice of what fits best for a given process.  The understanding of the different choices made by experiments contains important theoretical information about the SM and in the worst case scenario new physics could be inadvertently missed from being discovered. 

Another important lesson reemphasized by this study is the need for further theoretical investigations of jet vetos.  As we have shown \pt resummation causes a sizable effect on the total cross section because of the interplay between the jet veto and the \pt distribution.  Clearly the correlation demonstrated in Section~\ref{s.reweighting}, especially Fig. \ref{fig:JetVeto}, shows that the effects calculated in jet-veto resummation should be well approximated by the method employed here, similar to what was shown in~\cite{Banfi:2012yh}.  Of course there are additional logs related to the jet veto which cannot be systematically improved upon within \pt resummation.  It would be interesting to further investigate the interplay of these two types of resummation and the reweighting of parton showered events for more processes.

Another interesting question associated with the jet veto is how the LHC experiments can test the effects of the jet veto on the \ww cross section measurement.  The jet veto is a necessary evil in the context of measuring \ww without being overwhelmed by \ttb.  However, if the jet veto were weakened significantly, then the effects demonstrated in this note would disappear both in the context of \pt resummation and jet veto resummation.  If the jet-veto were varied, this could be compared to definitive predictions for the cross section as a function of the jet veto.  To alleviate the issue of the \ttb background, we suggest that the experiments separately implement a b-jet veto and a light jet veto, of which the light jet veto should be varied to study its effects.

In this paper we haven't explicitly demonstrated the effects of resummation on the contribution of $gg \rightarrow W^+W^-$ to the \ww cross section.  This contribution is a small fraction of the total cross section, and as such even though resummation effects will modify its shape as well, it won't change our conclusions.  However, it is important to note that the peak of the \pt spectrum for $gg \rightarrow W^+W^-$ should be at approximately 10 GeV high than for quark initiated \ww, as is generic for $gg$ initiated processes, in e.g. \cite{Bozzi:2005wk}).  For a sufficiently precise measurement of the \pt distribution it would be necessary to have the shape of this distribution correct as well.  A more interesting direction is the implications of understanding the correct shape of the SM \ww production background for the extraction of the Higgs signal in the $H\rightarrow W^+W^-$ decay channel.   Given that the \ww background is extracted via data-driven methods, it is important   
that the shape of the distributions of the \ww background is known when extrapolating from control to signal regions.  While the \pt of the \ww system isn't a variable used for the signal/control regions, as shown in our results for the reweighted kinematic distributions at 7 TeV there is a non negligible effect on the shape of relevant variables.  Future investigation is needed to study the effects of resummation on the measured signal strength of the Higgs in the \ww channel.

There are other avenues for future study, for instance investigating simultaneous resummation of \ww \pt with other observables, such as rapidity, to determine if any of the other cuts put on the fiducial phase space could alter the extraction of a total cross section.  Regardless of future direction, this work has clearly demonstrated the importance of \pt resummation when combined with fiducial phase space cuts.  Similar to how the \pt distribution of the Higgs signal is reweighted to make precise predictions for Higgs physics, it is important to use the correct \pt shape when considering processes where the \ww signal is either being measured or is an important background. To help facilitate future studies we plan to distribute the underlying \pt resummed distributions used in this study to any group interested in using them via a website.

\subsection*{Acknowledgements}
We would like to thank Karen Chen, David Curtin, Rafael Lopes de Sa, Dmytro Kovalskyi, Marc-Andre Pleier, Ted Rogers, and George Sterman for helpful discussions.
The work of P.M. was supported in part by NSF CAREER Award NSF-PHY-1056833.  The work of H.R and M.Z was supported in part by NSF grant PHY-1316617.
 
%%%%%%%%%%%%%%%%%%%%%%%%%%%%%%%%%%%%%%%%%%%%%%%%%%%%%%%%%%%%%%%%%%%%%%
%%%%%%%%%%%%%%%%%%%%%%%%%%%%%%%%%%%%%%%%%%%%%%%%%%%%%%%%%%%%%%%%%%%%%
%%%%%%%%%%%%%%%%%%%%%%%%%%%%%%%%%%%%%%%%%%%%%%%%%%%%%%%%%%%%%%%%%%%%%%
%%%%%%%%%%%%%%%%%%%%%%%%%%%%%%%%%%%%%%%%%%%%%%%%%%%%%%%%%%%%%%%%%%%%%


\begin{thebibliography}{99}

\bibitem{atlas7ww} 
  G.~Aad {\it et al.}  [ATLAS Collaboration],
  ``Measurement of $W^+W^-$ production in $pp$ collisions at $\sqrt{s}=7$ TeV with the ATLAS detector and limits on anomalous $WWZ$ and $W\gamma$ couplings,''
  arXiv:1210.2979 [hep-ex].
  %%CITATION = ARXIV:1210.2979;%%
  %11 citations counted in INSPIRE as of 25 Apr 2013

  \bibitem{atlas8ww}
  ``Measurement of $W^+W^-$ production in $pp$ collisions at $\sqrt{s}=8$ TeV with the ATLAS detector", ATLAS-CONF-2014-033

  
  \bibitem{cms7ww}
``Measurement of WW production rate" - CMS-PASSMP-12-005

  \bibitem{cms8ww}
``Measurement of WW production rate" - CMS-PASSMP-12-013


\bibitem{Chatrchyan:2013iaa} 
  S.~Chatrchyan {\it et al.}  [CMS Collaboration],
  %``Measurement of Higgs boson production and properties in the WW decay channel with leptonic final states,''
  JHEP {\bf 1401}, 096 (2014)
  [arXiv:1312.1129 [hep-ex]].
  %%CITATION = ARXIV:1312.1129;%%
  %37 citations counted in INSPIRE as of 15 Jul 2014

%\cite{Aad:2013wqa}
\bibitem{Aad:2013wqa} 
  G.~Aad {\it et al.}  [ATLAS Collaboration],
  %``Measurements of Higgs boson production and couplings in diboson final states with the ATLAS detector at the LHC,''
  Phys.\ Lett.\ B {\bf 726}, 88 (2013)
  [arXiv:1307.1427 [hep-ex]].
  %%CITATION = ARXIV:1307.1427;%%
  %204 citations counted in INSPIRE as of 15 Jul 2014

%\cite{Davatz:2004zg}
\bibitem{Davatz:2004zg} 
  G.~Davatz, G.~Dissertori, M.~Dittmar, M.~Grazzini and F.~Pauss,
  %``Effective K factors for gg ---> H ---> WW ---> l nu l nu at the LHC,''
  JHEP {\bf 0405}, 009 (2004)
  [hep-ph/0402218].
  %%CITATION = HEP-PH/0402218;%%
  %62 citations counted in INSPIRE as of 15 Jul 2014

\bibitem{Curtin:2012nn} 
  D.~Curtin, P.~Jaiswal and P.~Meade,
  %``Charginos Hiding In Plain Sight,''
  Phys.\ Rev.\ D {\bf 87}, no. 3, 031701 (2013)
  [arXiv:1206.6888 [hep-ph]].
  %%CITATION = ARXIV:1206.6888;%%
  %21 citations counted in INSPIRE as of 15 Jul 2014

%\cite{Curtin:2013gta}
\bibitem{Curtin:2013gta} 
  D.~Curtin, P.~Jaiswal, P.~Meade and P.~-J.~Tien,
  %``Casting Light on BSM Physics with SM Standard Candles,''
  JHEP {\bf 1308}, 068 (2013)
  [arXiv:1304.7011 [hep-ph]].
  %%CITATION = ARXIV:1304.7011;%%
  %5 citations counted in INSPIRE as of 15 Jul 2014

\bibitem{Rolbiecki:2013fia} 
  K.~Rolbiecki and K.~Sakurai,
  %``Light stops emerging in WW cross section measurements?,''
  JHEP {\bf 1309}, 004 (2013)
  [arXiv:1303.5696 [hep-ph]].
  %%CITATION = ARXIV:1303.5696;%%
  %8 citations counted in INSPIRE as of 15 Jul 2014

\bibitem{Curtin:2014zua} 
  D.~Curtin, P.~Meade and P.~-J.~Tien,
  %``Natural SUSY in Plain Sight,''
  arXiv:1406.0848 [hep-ph].
  %%CITATION = ARXIV:1406.0848;%%
  %2 citations counted in INSPIRE as of 15 Jul 2014


\bibitem{Kim:2014eva} 
  J.~S.~Kim, K.~Rolbiecki, K.~Sakurai and J.~Tattersall,
  %```Stop' that ambulance! New physics at the LHC?,''
  arXiv:1406.0858 [hep-ph].
  %%CITATION = ARXIV:1406.0858;%%
  %2 citations counted in INSPIRE as of 15 Jul 2014


\bibitem{Jaiswal:2013xra} 
  P.~Jaiswal, K.~Kopp and T.~Okui,
  %``Higgs Production Amidst the LHC Detector,''
  Phys.\ Rev.\ D {\bf 87}, no. 11, 115017 (2013)
  [arXiv:1303.1181 [hep-ph]].
  %%CITATION = ARXIV:1303.1181;%%
  %2 citations counted in INSPIRE as of 15 Jul 2014



\bibitem{Kauer:2012hd} 
  N.~Kauer and G.~Passarino,
  %``Inadequacy of zero-width approximation for a light Higgs boson signal,''
  JHEP {\bf 1208}, 116 (2012)
  [arXiv:1206.4803 [hep-ph]].
  %%CITATION = ARXIV:1206.4803;%%
  %68 citations counted in INSPIRE as of 14 Jul 2014

%\cite{Kauer:2013qba}
\bibitem{Kauer:2013qba} 
  N.~Kauer,
  %``Interference effects for H $\to$ WW/ZZ $\to \ell\bar{\nu}_\ell\bar{\ell}\nu_\ell$ searches in gluon fusion at the LHC,''
  JHEP {\bf 1312}, 082 (2013)
  [arXiv:1310.7011 [hep-ph]].
  %%CITATION = ARXIV:1310.7011;%%
  %7 citations counted in INSPIRE as of 14 Jul 2014

%\cite{Cascioli:2014yka}
\bibitem{Cascioli:2014yka} 
  F.~Cascioli, T.~Gehrmann, M.~Grazzini, S.~Kallweit, P.~Maierh歠er, A.~von Manteuffel, S.~Pozzorini and D.~Rathlev {\it et al.},
  ``ZZ production at hadron colliders in NNLO QCD,''
  arXiv:1405.2219 [hep-ph].
  %%CITATION = ARXIV:1405.2219;%%
  %7 citations counted in INSPIRE as of 10 Jul 2014
  
  %\cite{Dawson:2013lya}
\bibitem{Dawson:2013lya} 
  S.~Dawson, I.~M.~Lewis and M.~Zeng,
  %``Threshold resummed and approximate next-to-next-to-leading order results for $W^+W^-$ pair production at the LHC,''
  Phys.\ Rev.\ D {\bf 88}, no. 5, 054028 (2013)
  [arXiv:1307.3249].
  %%CITATION = ARXIV:1307.3249;%%
  %16 citations counted in INSPIRE as of 10 Jul 2014


%\cite{Magnea:1990zb}
\bibitem{Magnea:1990zb} 
  L.~Magnea and G.~F.~Sterman,
  %``Analytic continuation of the Sudakov form-factor in QCD,''
  Phys.\ Rev.\ D {\bf 42}, 4222 (1990).
  %%CITATION = PHRVA,D42,4222;%%
  %182 citations counted in INSPIRE as of 14 Jul 2014

%\cite{Ahrens:2008qu}
\bibitem{Ahrens:2008qu} 
  V.~Ahrens, T.~Becher, M.~Neubert and L.~L.~Yang,
  %``Origin of the Large Perturbative Corrections to Higgs Production at Hadron Colliders,''
  Phys.\ Rev.\ D {\bf 79}, 033013 (2009)
  [arXiv:0808.3008 [hep-ph]].
  %%CITATION = ARXIV:0808.3008;%%
  %79 citations counted in INSPIRE as of 13 Jul 2014

%\cite{Ahrens:2008nc}
\bibitem{Ahrens:2008nc} 
  V.~Ahrens, T.~Becher, M.~Neubert and L.~L.~Yang,
  %``Renormalization-Group Improved Prediction for Higgs Production at Hadron Colliders,''
  Eur.\ Phys.\ J.\ C {\bf 62}, 333 (2009)
  [arXiv:0809.4283 [hep-ph]].
  %%CITATION = ARXIV:0809.4283;%%
  %109 citations counted in INSPIRE as of 13 Jul 2014




%MC's, HqT, gg2VV-------------------------------------------------

%\cite{Kauer:2012hd}
\bibitem{Grazzini:2005vw} 
  M.~Grazzini,
  %``Soft-gluon effects in WW production at hadron colliders,''
  JHEP {\bf 0601}, 095 (2006)
  [hep-ph/0510337].
  %%CITATION = HEP-PH/0510337;%%
  %30 citations counted in INSPIRE as of 14 Jul 2014


\bibitem{Wang:2013qua} 
  Y.~Wang, C.~S.~Li, Z.~L.~Liu, D.~Y.~Shao and H.~T.~Li,
  %``Transverse-Momentum Resummation for Gauge Boson Pair Production at the Hadron Collider,''
  Phys.\ Rev.\ D {\bf 88}, 114017 (2013)
  [arXiv:1307.7520].
  %%CITATION = ARXIV:1307.7520;%%
  %4 citations counted in INSPIRE as of 10 Jul 2014

\bibitem{D0:2013jba} 
  V.~M.~Abazov {\it et al.}  [D0 Collaboration],
  %``Measurement of the $W$ boson mass with the D0 detector,''
  Phys.\ Rev.\ D {\bf 89}, no. 1, 012005 (2014)
  [arXiv:1310.8628 [hep-ex]].
  %%CITATION = ARXIV:1310.8628;%%
  %3 citations counted in INSPIRE as of 15 Jul 2014

%\cite{deFlorian:2011xf}
\bibitem{deFlorian:2011xf} 
  D.~de Florian, G.~Ferrera, M.~Grazzini and D.~Tommasini,
  %``Transverse-momentum resummation: Higgs boson production at the Tevatron and the LHC,''
  JHEP {\bf 1111}, 064 (2011)
  [arXiv:1109.2109 [hep-ph]].
  %%CITATION = ARXIV:1109.2109;%%
  %109 citations counted in INSPIRE as of 14 Jul 2014

%jet veto------------------------------------------------------

%\cite{Banfi:2012yh}
\bibitem{Banfi:2012yh} 
  A.~Banfi, G.~P.~Salam and G.~Zanderighi,
  %``NLL+NNLO predictions for jet-veto efficiencies in Higgs-boson and Drell-Yan production,''
  JHEP {\bf 1206}, 159 (2012)
  [arXiv:1203.5773 [hep-ph]].
  %%CITATION = ARXIV:1203.5773;%%
  %47 citations counted in INSPIRE as of 13 Jul 2014

%\cite{Banfi:2012jm}
\bibitem{Banfi:2012jm} 
  A.~Banfi, P.~F.~Monni, G.~P.~Salam and G.~Zanderighi,
  %``Higgs and Z-boson production with a jet veto,''
  Phys.\ Rev.\ Lett.\  {\bf 109}, 202001 (2012)
  [arXiv:1206.4998 [hep-ph]].
  %%CITATION = ARXIV:1206.4998;%%
  %43 citations counted in INSPIRE as of 13 Jul 2014

%\cite{Tackmann:2012bt}
\bibitem{Tackmann:2012bt} 
  F.~J.~Tackmann, J.~R.~Walsh and S.~Zuberi,
  %``Resummation Properties of Jet Vetoes at the LHC,''
  Phys.\ Rev.\ D {\bf 86}, 053011 (2012)
  [arXiv:1206.4312 [hep-ph]].
  %%CITATION = ARXIV:1206.4312;%%
  %26 citations counted in INSPIRE as of 13 Jul 2014

%\cite{Stewart:2013faa}
\bibitem{Stewart:2013faa} 
  I.~W.~Stewart, F.~J.~Tackmann, J.~R.~Walsh and S.~Zuberi,
  %``Jet $p_T$ Resummation in Higgs Production at $NNLL'+NNLO$,''
  Phys.\ Rev.\ D {\bf 89}, 054001 (2014)
  [arXiv:1307.1808].
  %%CITATION = ARXIV:1307.1808;%%
  %24 citations counted in INSPIRE as of 13 Jul 2014

%\cite{Becher:2012qa}
\bibitem{Becher:2012qa} 
  T.~Becher and M.~Neubert,
  %``Factorization and NNLL Resummation for Higgs Production with a Jet Veto,''
  JHEP {\bf 1207}, 108 (2012)
  [arXiv:1205.3806 [hep-ph]].
  %%CITATION = ARXIV:1205.3806;%%
  %41 citations counted in INSPIRE as of 13 Jul 2014

%\cite{Becher:2013xia}
\bibitem{Becher:2013xia} 
  T.~Becher, M.~Neubert and L.~Rothen,
  %``Factorization and $N^{3}LL_{p}$+NNLO predictions for the Higgs cross section with a jet veto,''
  JHEP {\bf 1310}, 125 (2013)
  [arXiv:1307.0025 [hep-ph]].
  %%CITATION = ARXIV:1307.0025;%%
  %22 citations counted in INSPIRE as of 13 Jul 2014

%\cite{Moult:2014pja}
\bibitem{Moult:2014pja} 
  I.~Moult and I.~W.~Stewart,
  %``Jet Vetoes Interfering with H->WW,''
  arXiv:1405.5534 [hep-ph].
  %%CITATION = ARXIV:1405.5534;%%
  %1 citations counted in INSPIRE as of 13 Jul 2014

% pi^2 ------------------------------------------------------

%\cite{Magnea:1990zb}
\bibitem{Dokshitzer:1978yd} 
  Y.~L.~Dokshitzer, D.~Diakonov and S.~I.~Troian,
  %``On the Transverse Momentum Distribution of Massive Lepton Pairs,''
  Phys.\ Lett.\ B {\bf 79}, 269 (1978).
  %%CITATION = PHLTA,B79,269;%%
  %176 citations counted in INSPIRE as of 06 May 2014

%\cite{Parisi:1979se}
\bibitem{Parisi:1979se} 
  G.~Parisi and R.~Petronzio,
  %``Small Transverse Momentum Distributions in Hard Processes,''
  Nucl.\ Phys.\ B {\bf 154}, 427 (1979).
  %%CITATION = NUPHA,B154,427;%%
  %476 citations counted in INSPIRE as of 06 May 2014

%\cite{Curci:1979bg}
\bibitem{Curci:1979bg} 
  G.~Curci, M.~Greco and Y.~Srivastava,
  %``{QCD} Jets From Coherent States,''
  Nucl.\ Phys.\ B {\bf 159}, 451 (1979).
  %%CITATION = NUPHA,B159,451;%%
  %145 citations counted in INSPIRE as of 06 May 2014

%\cite{Collins:1981uk}
\bibitem{Collins:1981uk} 
  J.~C.~Collins and D.~E.~Soper,
  %``Back-To-Back Jets in QCD,''
  Nucl.\ Phys.\ B {\bf 193}, 381 (1981)
  [Erratum-ibid.\ B {\bf 213}, 545 (1983)]
  [Nucl.\ Phys.\ B {\bf 213}, 545 (1983)].
  %%CITATION = NUPHA,B193,381;%%
  %826 citations counted in INSPIRE as of 06 May 2014

%\cite{Collins:1981va}
\bibitem{Collins:1981va} 
  J.~C.~Collins and D.~E.~Soper,
  %``Back-To-Back Jets: Fourier Transform from B to K-Transverse,''
  Nucl.\ Phys.\ B {\bf 197}, 446 (1982).
  %%CITATION = NUPHA,B197,446;%%
  %324 citations counted in INSPIRE as of 06 May 2014

%\cite{Kodaira:1981nh}
\bibitem{Kodaira:1981nh} 
  J.~Kodaira and L.~Trentadue,
  %``Summing Soft Emission in QCD,''
  Phys.\ Lett.\ B {\bf 112}, 66 (1982).
  %%CITATION = PHLTA,B112,66;%%
  %318 citations counted in INSPIRE as of 06 May 2014

%\cite{Kodaira:1982az}
\bibitem{Kodaira:1982az} 
  J.~Kodaira and L.~Trentadue,
  %``Single Logarithm Effects in electron-Positron Annihilation,''
  Phys.\ Lett.\ B {\bf 123}, 335 (1983).
  %%CITATION = PHLTA,B123,335;%%
  %131 citations counted in INSPIRE as of 06 May 2014

%\cite{Altarelli:1984pt}
\bibitem{Altarelli:1984pt} 
  G.~Altarelli, R.~K.~Ellis, M.~Greco and G.~Martinelli,
  %``Vector Boson Production at Colliders: A Theoretical Reappraisal,''
  Nucl.\ Phys.\ B {\bf 246}, 12 (1984).
  %%CITATION = NUPHA,B246,12;%%
  %506 citations counted in INSPIRE as of 06 May 2014

%\cite{Collins:1984kg}
\bibitem{Collins:1984kg} 
  J.~C.~Collins, D.~E.~Soper and G.~F.~Sterman,
  %``Transverse Momentum Distribution in Drell-Yan Pair and W and Z Boson Production,''
  Nucl.\ Phys.\ B {\bf 250}, 199 (1985).
  %%CITATION = NUPHA,B250,199;%%
  %699 citations counted in INSPIRE as of 06 May 2014

%\cite{Catani:2000vq}
\bibitem{Catani:2000vq} 
  S.~Catani, D.~de Florian and M.~Grazzini,
  %``Universality of nonleading logarithmic contributions in transverse momentum distributions,''
  Nucl.\ Phys.\ B {\bf 596}, 299 (2001)
  [hep-ph/0008184].
  %%CITATION = HEP-PH/0008184;%%
  %91 citations counted in INSPIRE as of 06 May 2014

%\cite{Bozzi:2005wk}
\bibitem{Bozzi:2005wk} 
  G.~Bozzi, S.~Catani, D.~de Florian and M.~Grazzini,
  %``Transverse-momentum resummation and the spectrum of the Higgs boson at the LHC,''
  Nucl.\ Phys.\ B {\bf 737}, 73 (2006)
  [hep-ph/0508068].
  %%CITATION = HEP-PH/0508068;%%
  %227 citations counted in INSPIRE as of 06 May 2014

%\cite{Bozzi:2010xn}
\bibitem{Bozzi:2010xn} 
  G.~Bozzi, S.~Catani, G.~Ferrera, D.~de Florian and M.~Grazzini,
  %``Production of Drell-Yan lepton pairs in hadron collisions: Transverse-momentum resummation at next-to-next-to-leading logarithmic accuracy,''
  Phys.\ Lett.\ B {\bf 696}, 207 (2011)
  [arXiv:1007.2351 [hep-ph]].
  %%CITATION = ARXIV:1007.2351;%%
  %44 citations counted in INSPIRE as of 06 May 2014

%\cite{deFlorian:2007sr}
\bibitem{deFlorian:2007sr} 
  D.~de Florian and J.~Zurita,
  %``Soft-gluon resummation for pseudoscalar Higgs boson production at hadron colliders,''
  Phys.\ Lett.\ B {\bf 659}, 813 (2008)
  [arXiv:0711.1916 [hep-ph]].
  %%CITATION = ARXIV:0711.1916;%%
  %2 citations counted in INSPIRE as of 06 May 2014

%\cite{Frixione:1993yp}
\bibitem{Frixione:1993yp} 
  S.~Frixione,
  %``A Next-to-leading order calculation of the cross-section for the production of W+ W- pairs in hadronic collisions,''
  Nucl.\ Phys.\ B {\bf 410}, 280 (1993).
  %%CITATION = NUPHA,B410,280;%%
  %115 citations counted in INSPIRE as of 07 May 2014
  
  
%\cite{Catani:2003zt}
\bibitem{Catani:2003zt} 
  S.~Catani, D.~de Florian, M.~Grazzini and P.~Nason,
  %``Soft gluon resummation for Higgs boson production at hadron colliders,''
  JHEP {\bf 0307}, 028 (2003)
  [hep-ph/0306211].
  %%CITATION = HEP-PH/0306211;%%
  %413 citations counted in INSPIRE as of 07 May 2014

\bibitem{Becher:2010tm} 
  T.~Becher and M.~Neubert,
  %``Drell-Yan production at small q_T, transverse parton distributions and the collinear anomaly,''
  Eur.\ Phys.\ J.\ C {\bf 71}, 1665 (2011)
  [arXiv:1007.4005 [hep-ph]].
  %%CITATION = ARXIV:1007.4005;%%
  %85 citations counted in INSPIRE as of 08 May 2014

%\cite{Martin:2009iq}
\bibitem{Martin:2009iq} 
  A.~D.~Martin, W.~J.~Stirling, R.~S.~Thorne and G.~Watt,
  %``Parton distributions for the LHC,''
  Eur.\ Phys.\ J.\ C {\bf 63}, 189 (2009)
  [arXiv:0901.0002 [hep-ph]].
  %%CITATION = ARXIV:0901.0002;%%
  %2057 citations counted in INSPIRE as of 07 May 2014

%\cite{Laenen:2000de}
\bibitem{Laenen:2000de} 
  E.~Laenen, G.~F.~Sterman and W.~Vogelsang,
  %``Higher order QCD corrections in prompt photon production,''
  Phys.\ Rev.\ Lett.\  {\bf 84}, 4296 (2000)
  [hep-ph/0002078].
  %%CITATION = HEP-PH/0002078;%%
  %112 citations counted in INSPIRE as of 07 May 2014

\bibitem{hqt} 
  G.~Bozzi, S.~Catani, D.~de Florian and M.~Grazzini,
  %``The q(T) spectrum of the Higgs boson at the LHC in QCD perturbation theory,''
  Phys.\ Lett.\ B {\bf 564}, 65 (2003)
  [hep-ph/0302104].
  %%CITATION = HEP-PH/0302104;%%
  %183 citations counted in INSPIRE as of 14 Jul 2014
    G.~Bozzi, S.~Catani, D.~de Florian and M.~Grazzini,
  %``Transverse-momentum resummation and the spectrum of the Higgs boson at the LHC,''
  Nucl.\ Phys.\ B {\bf 737}, 73 (2006)
  [hep-ph/0508068].
  %%CITATION = HEP-PH/0508068;%%
  %235 citations counted in INSPIRE as of 14 Jul 2014
    D.~de Florian, G.~Ferrera, M.~Grazzini and D.~Tommasini,
  %``Transverse-momentum resummation: Higgs boson production at the Tevatron and the LHC,''
  JHEP {\bf 1111}, 064 (2011)
  [arXiv:1109.2109 [hep-ph]].
  %%CITATION = ARXIV:1109.2109;%%
  %110 citations counted in INSPIRE as of 14 Jul 2014
  
  
  %\cite{Nason:2004rx}
\bibitem{Nason:2004rx} 
  P.~Nason,
  %``A New method for combining NLO QCD with shower Monte Carlo algorithms,''
  JHEP {\bf 0411}, 040 (2004)
  [hep-ph/0409146].
  %%CITATION = HEP-PH/0409146;%%
  %640 citations counted in INSPIRE as of 14 Jul 2014

%\cite{Frixione:2007vw}
\bibitem{Frixione:2007vw} 
  S.~Frixione, P.~Nason and C.~Oleari,
  %``Matching NLO QCD computations with Parton Shower simulations: the POWHEG method,''
  JHEP {\bf 0711}, 070 (2007)
  [arXiv:0709.2092 [hep-ph]].
  %%CITATION = ARXIV:0709.2092;%%
  %886 citations counted in INSPIRE as of 14 Jul 2014

%\cite{Alioli:2010xd}
\bibitem{Alioli:2010xd} 
  S.~Alioli, P.~Nason, C.~Oleari and E.~Re,
  %``A general framework for implementing NLO calculations in shower Monte Carlo programs: the POWHEG BOX,''
  JHEP {\bf 1006}, 043 (2010)
  [arXiv:1002.2581 [hep-ph]].
  %%CITATION = ARXIV:1002.2581;%%
  %477 citations counted in INSPIRE as of 14 Jul 2014

%\cite{Frixione:2002ik}
\bibitem{Frixione:2002ik} 
  S.~Frixione and B.~R.~Webber,
  %``Matching NLO QCD computations and parton shower simulations,''
  JHEP {\bf 0206}, 029 (2002)
  [hep-ph/0204244].
  %%CITATION = HEP-PH/0204244;%%
  %1582 citations counted in INSPIRE as of 14 Jul 2014
  
%\cite{Berger:2010xi}
\bibitem{Berger:2010xi} 
  C.~F.~Berger, C.~Marcantonini, I.~W.~Stewart, F.~J.~Tackmann and W.~J.~Waalewijn,
  %``Higgs Production with a Central Jet Veto at NNLL+NNLO,''
  JHEP {\bf 1104}, 092 (2011)
  [arXiv:1012.4480 [hep-ph]].
  %%CITATION = ARXIV:1012.4480;%%
  %67 citations counted in INSPIRE as of 04 Aug 2014

%\cite{Alwall:2014hca}
\bibitem{Alwall:2014hca} 
  J.~Alwall, R.~Frederix, S.~Frixione, V.~Hirschi, F.~Maltoni, O.~Mattelaer, H.~-S.~Shao and T.~Stelzer {\it et al.},
  %``The automated computation of tree-level and next-to-leading order differential cross sections, and their matching to parton shower simulations,''
  arXiv:1405.0301 [hep-ph].
  %%CITATION = ARXIV:1405.0301;%%
  %29 citations counted in INSPIRE as of 14 Jul 2014

%\cite{Bahr:2008pv}
\bibitem{Bahr:2008pv} 
  M.~Bahr, S.~Gieseke, M.~A.~Gigg, D.~Grellscheid, K.~Hamilton, O.~Latunde-Dada, S.~Platzer and P.~Richardson {\it et al.},
  %``Herwig++ Physics and Manual,''
  Eur.\ Phys.\ J.\ C {\bf 58}, 639 (2008)
  [arXiv:0803.0883 [hep-ph]].
  %%CITATION = ARXIV:0803.0883;%%
  %771 citations counted in INSPIRE as of 14 Jul 2014

%\cite{Sjostrand:2006za}
\bibitem{Sjostrand:2006za} 
  T.~Sjostrand, S.~Mrenna and P.~Z.~Skands,
  %``PYTHIA 6.4 Physics and Manual,''
  JHEP {\bf 0605}, 026 (2006)
  [hep-ph/0603175].
  %%CITATION = HEP-PH/0603175;%%
  %5319 citations counted in INSPIRE as of 14 Jul 2014

%\cite{deFavereau:2013fsa}
\bibitem{Pumplin:2002vw} 
  J.~Pumplin, D.~R.~Stump, J.~Huston, H.~L.~Lai, P.~M.~Nadolsky and W.~K.~Tung,
  %``New generation of parton distributions with uncertainties from global QCD analysis,''
  JHEP {\bf 0207}, 012 (2002)
  [hep-ph/0201195].
  %%CITATION = HEP-PH/0201195;%%
  %3940 citations counted in INSPIRE as of 15 Jul 2014

\bibitem{deFavereau:2013fsa} 
  J.~de Favereau {\it et al.}  [DELPHES 3 Collaboration],
  %``DELPHES 3, A modular framework for fast simulation of a generic collider experiment,''
  JHEP {\bf 1402}, 057 (2014)
  [arXiv:1307.6346 [hep-ex]].
  %%CITATION = ARXIV:1307.6346;%%
  %98 citations counted in INSPIRE as of 14 Jul 2014

%\cite{Bozzi:2008bb}
\bibitem{Bozzi:2008bb} 
  G.~Bozzi, S.~Catani, G.~Ferrera, D.~de Florian and M.~Grazzini,
  %``Transverse-momentum resummation: A Perturbative study of Z production at the Tevatron,''
  Nucl.\ Phys.\ B {\bf 815}, 174 (2009)
  [arXiv:0812.2862 [hep-ph]].
  %%CITATION = ARXIV:0812.2862;%%
  %28 citations counted in INSPIRE as of 15 Jul 2014
 
%\cite{Kulesza:2002rh}
%% \bibitem{Kulesza:2002rh} 
%%   A.~Kulesza, G.~F.~Sterman and W.~Vogelsang,
%%   %``Joint resummation in electroweak boson production,''
%%   Phys.\ Rev.\ D {\bf 66}, 014011 (2002)
%%   [hep-ph/0202251].
  %%CITATION = HEP-PH/0202251;%%
  %104 citations counted in INSPIRE as of 15 Jul 2014 
 
%\cite{Larkoski:2014tva}
%\bibitem{Larkoski:2014tva} 
  %A.~J.~Larkoski, I.~Moult and D.~Neill,
  %``Toward Multi-Differential Cross Sections: Measuring Two Angularities on a Single Jet,''
 % arXiv:1401.4458 [hep-ph].
  %%CITATION = ARXIV:1401.4458;%%
  %2 citations counted in INSPIRE as of 15 Jul 2014

%\cite{Khachatryan:2014dia}
\bibitem{Khachatryan:2014dia} 
  V.~Khachatryan {\it et al.}  [CMS Collaboration],
  %``Measurement of the pp to ZZ production cross section and constraints on anomalous triple gauge couplings in four-lepton final states at $\sqrt{s}$ = 8 TeV,''
  arXiv:1406.0113 [hep-ex].
  %%CITATION = ARXIV:1406.0113;%%

%\cite{Pumplin:2002vw}

%\cite{Cascioli:2013gfa}
\bibitem{Cascioli:2013gfa} 
  F.~Cascioli, S.~Hoeche, F.~Krauss, P.~Maierh鰂er, S.~Pozzorini and F.~Siegert,
  %``Precise Higgs-background predictions: merging NLO QCD and squared quark-loop corrections to four-lepton + 0,1 jet production,''
  JHEP {\bf 1401}, 046 (2014)
  [arXiv:1309.0500 [hep-ph]].
  %%CITATION = ARXIV:1309.0500;%%
  %17 citations counted in INSPIRE as of 04 Aug 2014

%\cite{Campanario:2013wta}
\bibitem{Campanario:2013wta} 
  F.~Campanario, M.~Rauch and S.~Sapeta,
  %``$W^+W^-$ production at high transverse momenta beyond NLO,''
  Nucl.\ Phys.\ B {\bf 879}, 65 (2014)
  [arXiv:1309.7293 [hep-ph]].
  %%CITATION = ARXIV:1309.7293;%%
  %6 citations counted in INSPIRE as of 05 Aug 2014

%\cite{Campanario:2012fk}
\bibitem{Campanario:2012fk} 
  F.~Campanario and S.~Sapeta,
  %``WZ production beyond NLO for high-pT observables,''
  Phys.\ Lett.\ B {\bf 718}, 100 (2012)
  [arXiv:1209.4595 [hep-ph]].
  %%CITATION = ARXIV:1209.4595;%%
  %18 citations counted in INSPIRE as of 05 Aug 2014
  
  %\cite{Ohnemus:1991kk}
\bibitem{Ohnemus:1991kk} 
  J.~Ohnemus,
  %``An Order $\alpha^- s$ calculation of hadronic $W^{-} W^{+}$ production,''
  Phys.\ Rev.\ D {\bf 44}, 1403 (1991).
  %%CITATION = PHRVA,D44,1403;%%
  %185 citations counted in INSPIRE as of 11 Aug 2014

%\cite{Bonvini:2013jha}
\bibitem{Bonvini:2013jha} 
  M.~Bonvini, F.~Caola, S.~Forte, K.~Melnikov and G.~Ridolfi,
  %``Signal-background interference effects for $gg→H→W^+W^-$ beyond leading order,''
  Phys.\ Rev.\ D {\bf 88}, no. 3, 034032 (2013)
  [arXiv:1304.3053 [hep-ph]].
  %%CITATION = ARXIV:1304.3053;%%
  %19 citations counted in INSPIRE as of 18 Aug 2014

%\cite{Vogt:2004ns}
\bibitem{Vogt:2004ns} 
  A.~Vogt,
  %``Efficient evolution of unpolarized and polarized parton distributions with QCD-PEGASUS,''
  Comput.\ Phys.\ Commun.\  {\bf 170}, 65 (2005)
  [hep-ph/0408244].
  %%CITATION = HEP-PH/0408244;%%
  %108 citations counted in INSPIRE as of 18 Aug 2014

 \end{thebibliography}
\end{document}